# Bose-Einstein Condensation of Quasi-Particles by Rapid Cooling

**Authors:** M. Schneider[1], T. Brächer[1], D. Breitbach[1], V. Lauer[1], P. Pirro[1], D. A. Bozhko[2], H. Yu. Musiienko-Shmarova[1], B. Heinz[1,3], Q. Wang[1], T. Meyer[1,4], F. Heussner[1], S. Keller[1], E. Th. Papaioannou[1], B. Lägel[5], T. Löber[5], C. Dubs[6], A. N. Slavin[7], V. S. Tiberkevich[7], A. A. Serga[1], B. Hillebrands[1], and A. V. Chumak[1,8,*]

The fundamental phenomenon of Bose-Einstein Condensation (BEC) has been observed in different systems of real and quasi-particles. The condensation of real particles is achieved through a major reduction in temperature while for quasi-particles a mechanism of external injection of bosons by irradiation is required. Here, we present a novel and universal approach to enable BEC of quasi-particles and to corroborate it experimentally by using magnons as the Bose-particle model system. The critical point to this approach is the introduction of a disequilibrium of magnons with the phonon bath. After heating to an elevated temperature, a sudden decrease in the temperature of the phonons, which is approximately instant on the time scales of the magnon system, results in a large excess of incoherent magnons. The consequent spectral redistribution of these magnons triggers the Bose-Einstein condensation.

**State of the art of Bose-Einstein condensation.** Bosons are particles of integer spin that allow for the fundamental quantum effect of Bose–Einstein Condensation (BEC), which manifests itself in the formation of a macroscopic coherent state in an otherwise incoherent, thermalized many-particle system. The phenomenon of BEC was originally predicted for an ideal gas by Albert Einstein in 1924 based on the theory developed by Satyendra Nath Bose. Nowadays, Bose-Einstein condensates are investigated experimentally in a variety of different systems, which includes real particles such as ultra-cold gases[1,2], as well as quasi-particles with the likes of exciton–polaritons[3,4], photons[5,6], or magnons in quantum magnets[7-9], liquid helium $^{3}$H[10,11], and ferrimagnets[12-15]. The phenomenon can be reached by a major decrease in the system temperature or by an increase in the particle density. In order to condensate atomic gases, extremely low temperatures on the order of mK are required since the density of such gases must be very low to prevent their cohesion. In contrast, cooling of a quasi-particle system is accompanied by a decrease in its population and prevents BEC. Thus, an injection of bosons is required to reach the threshold for BEC in such systems. Since quasi-particle systems allow for high densities of bosons, BEC by quasi-particle injection is possible even at room temperature. Prominent examples are the injection of exciton-polaritons by a laser[3,4] and of magnons by either microwave parametric pumping[12–16] or due to the spin Seebeck effect induced spin currents[17].

**Concept of BEC by rapid cooling.** In this letter, we propose and demonstrate experimentally a different and universal way to achieve BEC of quasi-particles. In a solid body, each quasi-particle system interacts with the phonon bath and these two systems stay in equilibrium in a quasi-stationary state. An instant reduction in the phonon temperature results in a disequilibrium between these systems and in a large excess of quasi-particles when compared to the equilibrium state at the new temperature. In contrast to all previous studies[3-6,12-15], this kind of "injection" is a-priori incoherent, populates the entire energy spectrum rather than a narrow spectral range, and, consequently, excludes the necessity of an initial thermalization of the injected quasi-particles. The consequent redistribution of the quasi-particles due to multi-particle and particle-phonon scattering processes results in the increase in the chemical potential $\mu$ required for the BEC.

The energy-distribution $n\,(hf,\ \mu,\ T)$ of the density of bosons depends on temperature $T$ and is described by the Bose-Einstein distribution function multiplied with their energy-dependent density of states $D\,(hf)$:

$$n\left(hf,\mu(t),T(t)\right)=\frac{D\left(hf\right)}{\exp\left(\dfrac{hf-\mu(t)}{k_{\mathrm{B}}T(t)}\right)-1}\,, \tag{1}$$

where $hf$ is the energy of the quasi-particle, $f$ is its frequency, $t$ is the elapsed time, $T(t)$ is the phonon or lattice temperature, $\mu(t)$ is the chemical potential, $h$ is Planck's constant, and $k_{\mathrm{B}}$ is Boltzmann's constant. In the work presented, BEC is studied experimentally using a magnon system. For their description, the approximated density of states for ferromagnetic exchange magnons is used: $D\left(hf\right)\propto\sqrt{f-f_{\min}}$, where $f_{\min}$ is the frequency at the bottom of the magnon spectrum. The dashed blue lines in all panels of Fig. 1b show the steady-state density distributions when the magnon and the phonon systems are in equilibrium at room temperature $T=T_{\mathrm{Room}}$ for $\mu=0$. In order to achieve a BEC of magnons, the chemical potential $\mu$ has to be increased up to the value of the minimum magnon energy $hf_{\min}$. In this case, Equation (1) diverges, reflecting the condensation of the quasi particles into the same quantum state at the smallest energy of the system. The distribution of magnon density in the particular case of $\mu=hf_{\min}$ at room temperature $T=T_{\mathrm{Room}}$ is shown in Fig. 1b by the dashed green lines.

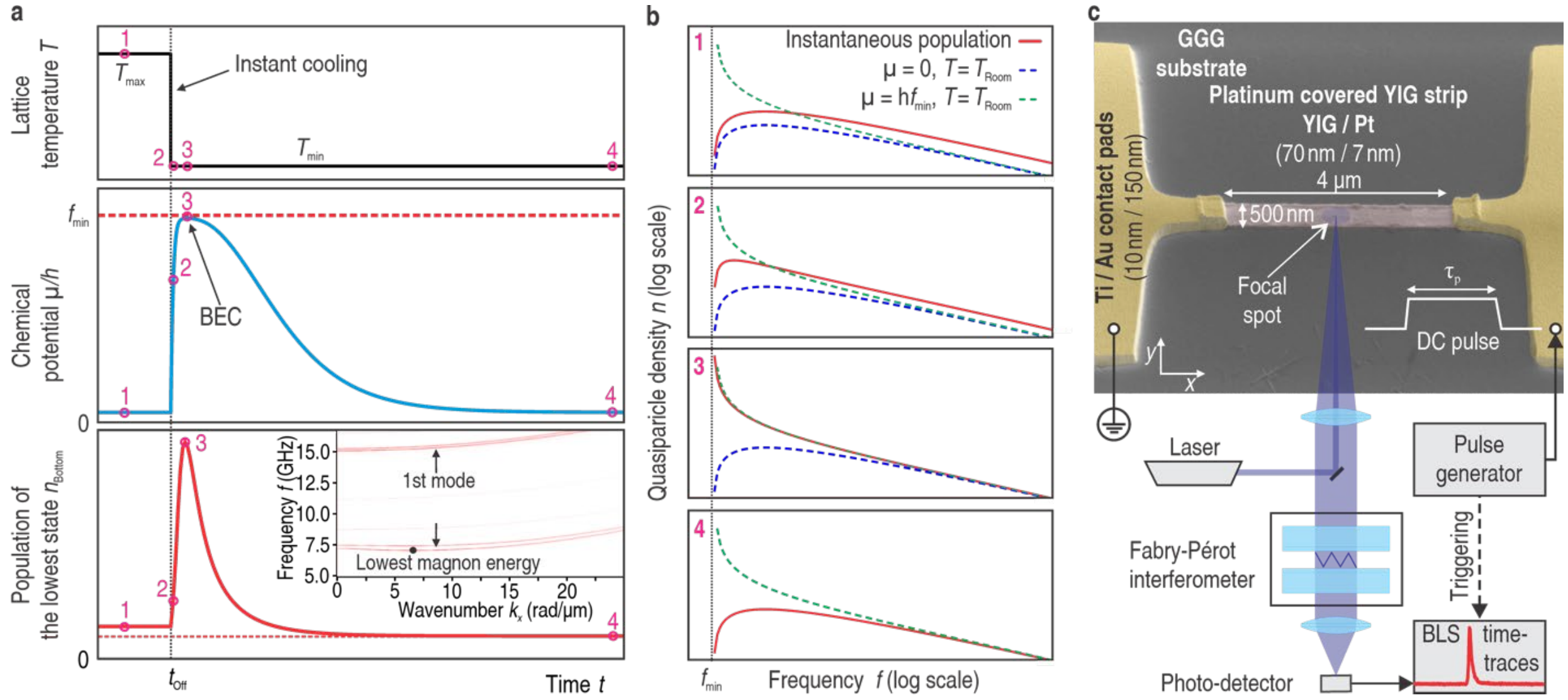

**Fig. 1 | Theoretical modelling of the BEC by rapid cooling and experimental approach. a**, and **b**, show schematic depiction of the condensation process. **a**, shows the time evolution of the temperature of the phonon system (top panel), the quasi-particle chemical potential (middle panel) and the quasi-particle population at the lowest energy state (bottom panel). The time of the instant cooling is marked by $t_{\mathrm{Off}}$. The inset shows the simulated magnon spectrum (white-to-red linear scale) for the YIG structure depicted in **c,** and magnetized along its long axis by an external field of $\mu_0 H_{\mathrm{ext}}=188$ mT. **b**, shows the quasi-particle density as a function of frequency. The red lines in the different panels show the densities for the different times marked in **a**. Dashed blue lines: Steady-state room-temperature distributions. Dashed green lines: Room-temperature distribution with $\mu=hf_{\min}$. **c,** Structure under investigation and sketch of the experimental setup.

Our approach to provide conditions for the BEC is as follows. First, the temperature of the phonon system is raised to a certain critical value. This rise in temperature increases the number of phonons and magnons simultaneously (see point “1” in Fig. 1a). The magnon density in the heated equilibrium state is given by Eq. 1 at the increased temperature and $\mu = 0$ is depicted in the Panel “1” in Fig. 1b by the solid red line. The proposed mechanism for BEC is based on the rapid cooling of the phonon system as it is shown in Fig. 1a. As the lattice temperature drops to $T_{\min}$, a disequilibrium between the magnon and phonon systems results in an excess of magnons over the whole spectrum with respect to the equilibrated state. The consequent magnon distribution (see Panel “2” in Fig. 1b) is calculated following the dynamic rate equations given in the Methods section. Two mechanisms are of highest importance for this magnon redistribution. The first one, is the phonon-magnon coupling that transfers energy from the magnon to the phonon system. With the assumption of the simplest case of viscous Gilbert damping[18], the rate of decay of the magnons is proportional to their energy. At the same time, magnon-magnon scattering processes[18], i.e., the interactions within the magnon system, allow for a redistribution of magnons towards low energies. Consequently, the density of low-energy magnons increases while the high-energy magnons dissipate. This leads to a decrease in the temperature and to a simultaneous increase in the chemical potential $\mu$ of magnons over time – see middle panel in Fig. 1a. If the critical number of magnons in the heated state is reached, the chemical potential will rise up to the minimum magnon energy $hf_{\min}$ and BEC of magnons occurs (see Panels “2” and “3” in Fig. 1b). This is manifested in the form of a sharp increase in the magnon density at the lowest energy state shown in the bottom panel of Fig. 1a. The simulated magnon spectrum (see Methods) is shown in the inset of Fig. 1a and the lowest magnon energy state is indicated. Ultimately, the magnon system returns to room-temperature equilibrium (see point “4” in Figs. 1a and Panel “4” in Fig. 1b) due to scattering into the phonon system, which is responsible for the finite magnon lifetime.

For simplicity, our model only considers the case of quasi-particle conserving four magnon scattering processes (see Methods) but, other mechanisms like Cherenkov radiation will also contribute to the magnon redistribution towards low energies[18,19] and can, thus, significantly enhance the condensation efficiency.

The proposed mechanism can be used universally for all bosonic quasi-particles showing internal inelastic scattering mechanisms and that are interacting with the phonon system in a way that the relaxation rate increases with the increase in the quasi-particle energy. The key experimental challenge, which has inhibited its realization so far, is the achievement of a sufficiently high cooling rate. Recent progress in the creation of (magnetic) nano-structures enabled quasi-instant cooling. In these structures, the phonon transport from a heated nano-structure to the quasi-bulk surroundings ensures a cooling time shorter than the characteristic magnon scattering time and the characteristic interaction time of the quasi-particles with the phonon bath.

**Experimental observation and proof of BEC by rapid cooling.** In our study, as model system, we investigate magnons at room temperature in an Yttrium Iron Garnet (YIG)[20] – Platinum nano-strip (width 500 nm, length 5 μm, YIG thickness 70 nm, Pt thickness 7 nm (see Fig. 1c). A biasing magnetic field sufficiently large to magnetize the strip along the field direction is applied either in-plane parallel (**B**||**x**) or perpendicular (**B**||**y**) to the long axis of the strip – for the exact geometry, see Fig. 1c. Ti/Au leads have been deposited on top of the strip to apply electric current pulses to the Pt layer, which heat up the YIG/Pt structure due to Joule heating. Switching off the pulse results in a fast cooling with rates on the order of tens of Kelvin per nanosecond.

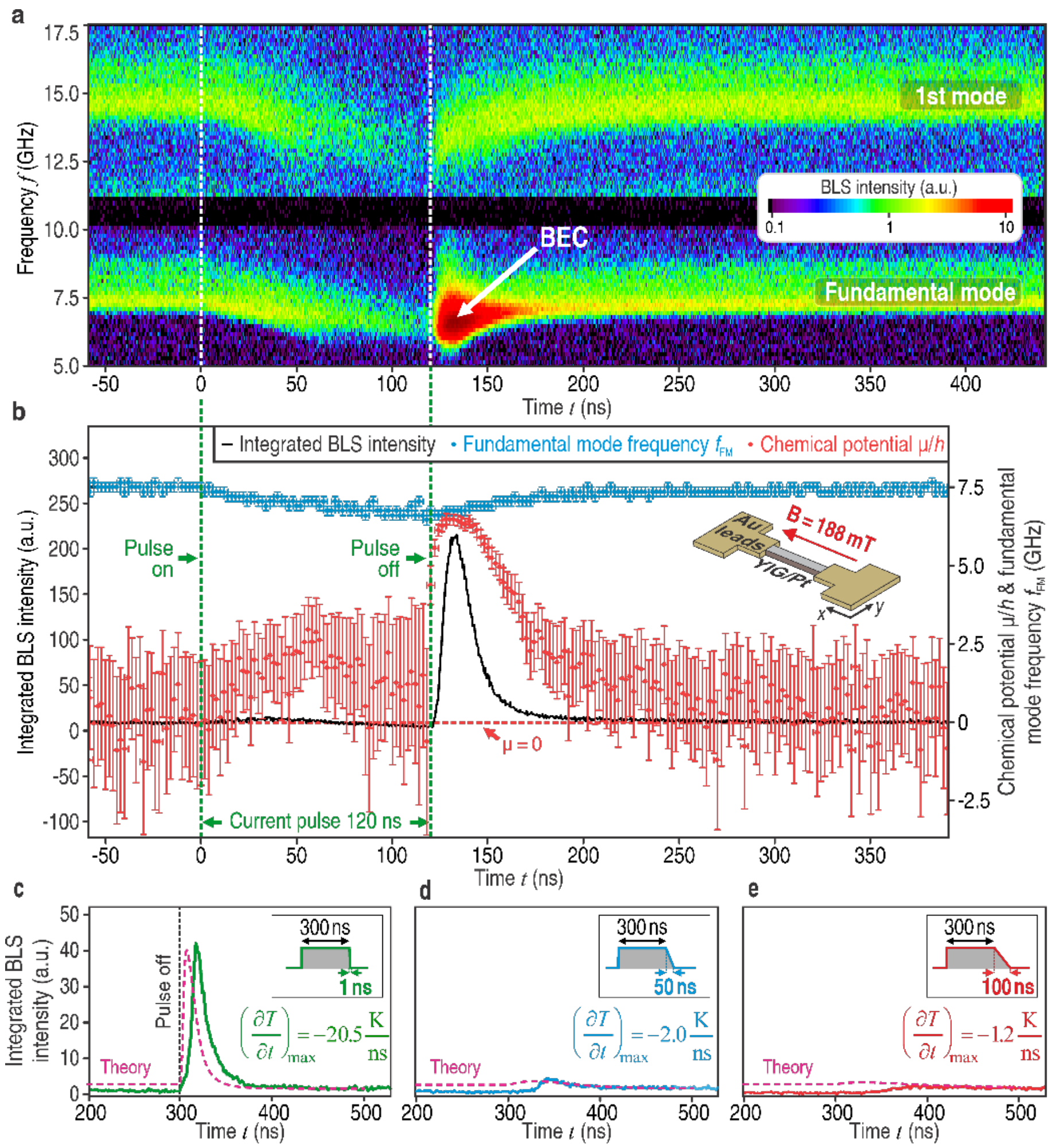

**Fig. 2 | BLS Measurements of the BEC by rapid cooling. a**, Depicts the measured BLS spectrum measured as a function of time for **B**||**x**. The BLS signal (color-coded, log scale) is proportional to the density of magnons. The vertical dashed lines indicate the start and end of the pulse. Switching off the pulse results in the formation of a pronounced magnon signal at the bottom of the magnon band. **b,** Normalized BLS intensity integrated from 4.95 GHz to 8.1 GHz as a function of time (black line), frequency of the fundamental mode (blue dots) and measured chemical potential (red dots). The BEC of magnons manifests itself as the peak in the BLS intensity that is proportional to the magnon density. Moreover, it is clearly visible that the chemical potential reaches the frequency of the fundamental mode. The inset shows the sample geometry and the coordinate system used. **C,** The magnon intensity at the bottom of the spectrum (analogue to panel b but with and **B**||**y**). The pink dashed line shows the magnon population at the bottom of the spectrum calculated using the dynamic rate equations discussed in the Methods Section with $\tau_{Fall} = 1$ ns. **D,** The same experiment with increased fall times of 50 ns (blue line). Here, the BEC is strongly suppressed. The pink dashed line shows the calculated magnon density for $\tau_{Fall} = 50$ ns. For this slow cooling, the chemical potential reaches a maximal value of only $0.58 \times f_{min}$. **e,** The same experiment with increased fall times of 100 ns (red line) and theory for $\tau_{Fall} = 100$ ns (pink line). See the inset for the time profiles of the pulses.

This fast cooling is provided by the phonon transport to the quasi-bulk Gadolinium Gallium Garnet (GGG) substrate and to the Ti/Au leads. In order to measure the time evolution of the magnon density in the YIG strip, micro-focused time-resolved Brillouin Light Scattering (BLS) spectroscopy[21] is used (see Methods).

Figure 2a shows the measured magnon spectrum as a function of time when the voltage pulse with a duration of $\tau_{\mathrm{P}} = 120\,\mathrm{ns}$ and a rise and fall time of $\tau_{\mathrm{Rise}} = \tau_{\mathrm{Fall}} = 1\,\mathrm{ns}$ is applied. The beginning and the end of the applied pulse are indicated by the vertical dashed lines. The amplitude of the voltage pulse was $U = 0.9$ V, corresponding to an estimated current density of $7.8\times10^{11}$ A/m$^2$ in the Pt overlayer. The in-plane biasing magnetic field of $\mu_0 H_{\mathrm{ext}} = 188$ mT is applied along the strip. The thermal population of two magnon modes is visible before the current pulse is applied. The magnon mode at $f_{\mathrm{FM}} = 7.5$ GHz has the smallest frequency and is the fundamental mode. The magnon mode of frequency $f_{\mathrm{1st\ mode}} = 14.9$ GHz is the first standing thickness mode[18]. Once the current pulse is switched on at the time $t = 0$ ns, the frequencies of both modes decrease with time due to heating and the consequent decrease in the saturation magnetization $M_{\mathrm{s}}$ and the exchange stiffness $D_{\mathrm{ex}}$ of YIG[22]. Another observable effect is the decrease of the BLS intensity during the current pulse that is due to a temperature-dependent decrease in the sensitivity of the BLS[23].

One can clearly see in Fig. 2a that switching off the current pulse at the time $t = \tau_{\mathrm{P}}$ results in the formation of a pronounced magnon signal at the frequency of the fundamental mode, which corresponds to the bottom of the magnon spectrum. This is the characteristic fingerprint of the process of BEC of magnons described by the proposed theoretical model. The signal maximum is reached approximately 10 ns after the current pulse is switched off. Afterwards, the BLS intensity decreases exponentially, corresponding to a magnon lifetime of $21 \pm 4$ ns. In addition, the recovery of the frequency of both modes is observed. The black line in Fig. 2b presents the time evolution of the BLS intensity integrated over the frequency range at the bottom of the magnon spectrum. A clear peak in the magnon density is in agreement with the result of the simplified theoretical model in the bottom panel of Fig. 1a.

The criterion of a BEC is that the chemical potential reaches the minimal energy of the quasi-particles. Our experiment allows for the direct measurement of the temporal evolution of the chemical potential $\mu/h$ since the BLS intensity is proportional to the magnon density and the density of the first standing mode can be used as a reference (see Methods). The red dots in Fig. 2b shows the measured chemical potential in GHz units. One can clearly see that the potential strongly increases after the current pulse is switched off and reaches the frequency of the fundamental mode defining the minimal magnon energy. This experimental finding confirms the BEC of magnons directly. It is also visible that the thermal equilibrium between the magnon and phonon bathes in the nanostructured YIG film is recovered on a time scale of about 100–150 ns, when the chemical potential of the magnon system recovers to its zero value. This time scale is consistent with the BEC process since it is sufficiently long for the BEC to be formed. It is remarkable that the temporal behaviors of the experimentally determined chemical potential agrees both qualitatively and quantitatively with the results of our theoretical calculations presented in the Extended Data Fig. 1.

Please note that a weak increase in the chemical potential is also observed during the current pulse. It can be attributed to the spin Seebeck effect (SSE), as reported recently[17,24,25]. From the simulated temperature gradient (to $3.5 \times 10^8$ K/m, see Supplement) and with the approximation that the spin-diffusion length is determined by the thickness of the YIG layer, the calculated[26] initial increase in the chemical potential caused by the spin Seebeck effect is of $\mu/h = 0.7 f_{\mathrm{min}} =$ 5.25 GHz. As the YIG sample heats up from room temperature to the threshold temperature of 440 K (see, Extended Data Fig. 5), this value, which is still not enough to reach the BEC in our experiment, rapidly decreases due to the tenfold reduction in the SSE magnitude (see Fig. 3c in Ref. [27]). As a result, no increase in the low-energy magnon density is observed during the application of the heating electric current even in the continuous regime. Thus, we can state that in contrast to the SSE-induced magnon BEC measured at low temperatures[17], where the maximal temperature of the YIG-layer was near room temperature, the thermally induced spin injection from the Pt layer plays only a minor role in our experiments. This statement is directly confirmed by our observations of magnon accumulation at $f_{\mathrm{min}}$ in an experiment with test structures containing an Al/Au heater instead of the Pt layer (see Supplement).

In addition, the same experiment was performed when the field was applied perpendicular to the long axis of the strip (**B**||**y**, see Supplement) in different directions. No qualitative difference between the results was observed excluding the spin Hall effect induced spin transfer torque[28-30] as the mechanism responsible for the experimental findings.

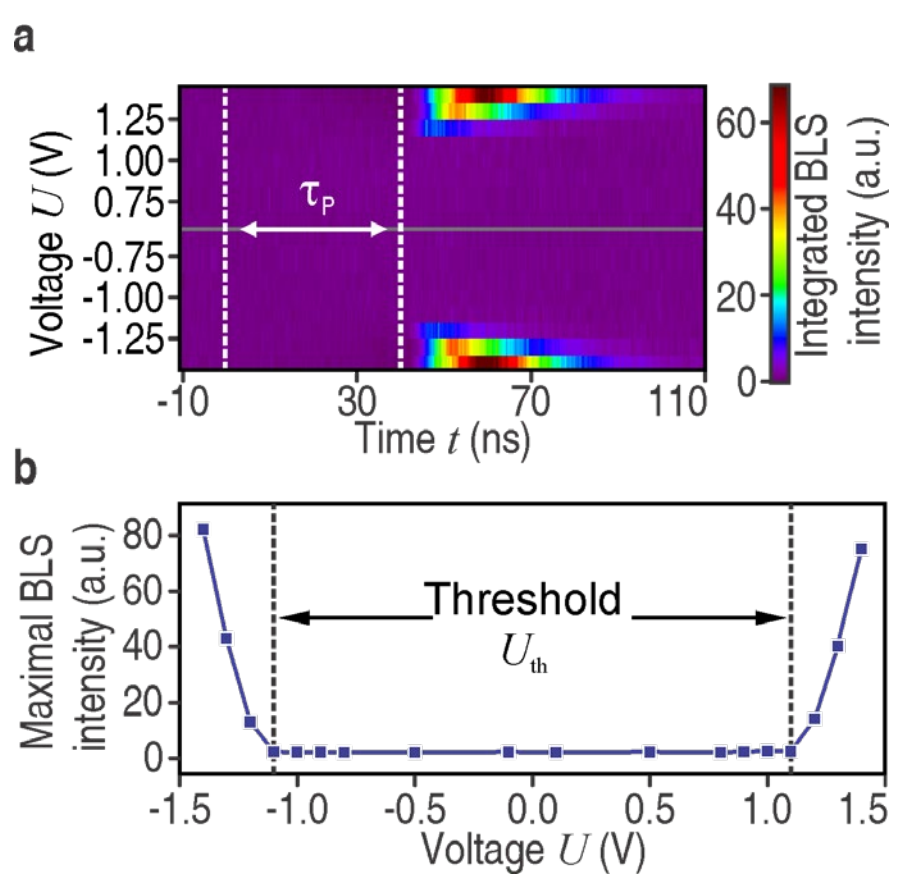

**Fig. 3 | Threshold behaviour of the BEC.** **a,** Measured intensity of magnons at the bottom of the spectrum (color-coded) as a function of time and applied voltage (**B**||**y**). The applied current pulse is marked by the vertical dashed lines. **b,** Maximal magnon intensity as a function of the applied voltage extracted from **a**. A clear threshold is visible at 1.2 V, independent of the current polarity.

To verify the crucial role of the fast cooling for the BEC, additional measurements in which the current pulses were switched off with longer fall times of 50 ns and 100 ns were performed. For these measurements, the pulse duration was increased to $\tau_{\mathrm{P}} = 300\,\mathrm{ns}$. As can be seen from Fig. 2c, the increased pulse duration does not inhibit the BEC. The influence of the fall time is shown in Figs. 2d, and 2e. It is evident that the condensate disappears for long fall times, i.e., slow cooling rates. The maximal temperature approaches a value of $T_{\mathrm{sim}}^{\mathrm{max}} = 491\,\mathrm{K}$ at the end of the current pulse according to simulations performed with COMSOL Multiphysics (see Supplement). The maximal cooling rates of the phonon system $\left(\partial T / \partial t\right)_{\mathrm{max}}$ obtained from these simulations are also indicated in the Figures. The values clearly demonstrate that the cooling rate of 2 K/ns is already too slow to trigger the magnon BEC, while a rate of 20.5 K/ns is still fast enough. Moreover, the simulations show that the cooling rate is not constant and decreases with time. In order to model this, the simple approach of an exponential decay $T \propto \exp\left(-t / \tau_{\mathrm{Fall}}\right)$ of the temperature was used in the theoretical model instead of the step function discussed in Fig. 1. The theoretical results are presented in Fig. 2c-e and support the experimental findings – the BEC is observed only in the first case of $\tau_{\mathrm{Fall}} = 1\,\mathrm{ns}$. Please note that the theoretical calculations are normalized to the intensity level of the thermal waves in equilibrium. Thus, also the generated total number of magnons is in good agreement with the predictions from the model.

The BEC of quasi-particles is a threshold-like phenomenon that takes place when the chemical potential μ reaches the minimal energy level $hf_{\mathrm{min}}$. This implies that in this experiment, a certain threshold temperature needs to be exceeded which corresponds to a certain critical magnon density in the system. Therefore, for a given pulse duration $\tau_{\mathrm{P}}$, a certain voltage needs to be applied to reach this critical density. The clear threshold-like behaviour is visible in Fig. 3a, in which the frequency-integrated BLS intensity (color-coded) is plotted as a function of time and voltage for $\tau_{\mathrm{P}} = 40\,\mathrm{ns}$. The extracted maximum magnon intensity is depicted in Fig. 3b as a function of the applied voltage. It can be clearly seen that the application of voltage pulses with amplitudes smaller than $\pm 1.2\,\mathrm{V}$ does not result in BEC, since the critical magnon density is not reached. The threshold magnon density is exceeded and the magnon condensate is formed only for larger voltages. Further details on the threshold nature can be found in the Supplement.

In general, the threshold behavior in Fig. 3 is one of the main three indicators of the BEC presented in our work. Together with the observation of spontaneous condensation of magnons at the lowest energy state (Fig. 2a) and with the experimentally determined increase in the chemical potential up to the energy of this state (Fig. 2b), it allows us to confidently declare the creation of the magnon Bose-Einstein condensate in our experiment.

**Conclusions.** In conclusion, a novel way to create a magnon Bose-Einstein condensate by rapid cooling in an individual magnetic nano-structure has been proposed and verified experimentally. BEC of magnons has been achieved in a typical spintronic structure by the application of mere current pulses with powers of a few mW rather than by the usage of (intricate) high-power microwave pumping. This paves the way for the usage of macroscopic quantum magnon states in conventional spintronics and to on-chip solid-state quantum computing. We would like to stress that the observed way to BEC is genuine to any solid-state quasi-particles in exchange with the phonon bath. The injection mechanism is originally incoherent, which is in direct opposition to laser or microwave irradiation and can be applied to other bosonic systems such as exciton-polaritons and photons in cavities.

# Methods

**Theoretical model of the magnon spectral redistribution process.** The spectral redistribution of a magnon gas in response to rapid changes of the lattice temperature can be modelled by the following model. For simplicity, we assume that the magnon distribution remains isotropic and depends only on the magnon frequency $f$, i.e. it is described by the distribution function $n(f,t)$. In the quasi-equilibrium state with lattice temperature $T_L(t)$ and chemical potential $\mu(t)$, the distribution function is given by Eq. (1) in the manuscript. The main reason we may approximate the real quantized spectrum as a continuous one is that, in our experiments and simulations, the magnon subsystem is heated and cooled by thermal contact with lattice and this process involves magnons in the whole thermal range – up to several THz. The frequency bandwidth of such magnons (which are primarily responsible for the nonlinear thermalization in the magnon subsystem) is much larger than the frequency distance between quantized magnon modes and, therefore, the magnon spectrum can be safely approximated as a continuum.

Numerical calculations were performed on a uniform frequency grid:

$$f_k = f_{\min} + \Delta f\, k \quad (2)$$

where $k$ is the step number and $\Delta f$ is the grid cell size. Now, instead of the continuous distribution function $n(f,t)$ we obtain a discrete set of magnon population numbers:

$$n_k(t) = n(f_k, t).$$

The dynamics of the population numbers $n_k(t)$ can be described by the following set of equations:

$$D_k\left[\frac{dn_k}{dt} + \Gamma_k\left(n_k - n_k^0(t)\right)\right] = N_k\ , \quad (3)$$

where $D_k = D(f_k)$is the density of states for the $k$-th magnon level ($D_k n_k$ is the density of magnons in the frequency range $f_k \pm \Delta f/2$). The density of states is determined by the dispersion law of quasi-particles and the dimensionality of space. In our case, for a quadratic dispersion in the 3D case[31]:

$$D_k(f_k) = \Delta f\sqrt{f_k - f_{\min}} \quad (4)$$

For the lowest state $k = 0$, $D_0 = 1/V$ was taken, where $V$ is the sample's volume. $\Gamma_k = \Gamma(f_k)$ in Eq. (3) is the spin-lattice relaxation time, for which one can use the simple "Gilbert" approximation of viscous damping[32]

$$\Gamma(f) = 2\alpha_G f. \quad (5)$$

$n_k^0(t) = n_0(f_k, t)$ in Eq. (3) is the instantaneous equilibrium magnon population, which is determined by the instantaneous lattice temperature $T_L(t)$

$$n^0(f,t) = n_{\text{eq}}\left(f, T_L(t), \mu = 0\right). \quad (6)$$

The right-hand side of Eq. (3) $N_k$ describes magnon transitions due to nonlinear four-magnon processes[18]. We consider the simplest model of nonlinear magnon scattering, in which only the scattering of magnons with similar energies was taken into account. Namely, we considered only the scattering processes of the form

$$f_k + f_k \leftrightarrow f_{k+1} + f_{k-1}\ . \quad (7)$$

The net rate of such processes is

$$F_k = C_k\left[n_k^2\left(n_{k+1}+1\right)\left(n_{k-1}+1\right)-n_{k+1}n_{k-1}\left(n_k+1\right)^2\right], \quad (8)$$

where $C_k = C\sqrt{f_k - f_{\min}}\left(\frac{f_k - f_{\min}}{\Delta f}\right)^3$. Here the constant $C$ is a fitting parameter, which defines the efficiency of the four magnon scattering processes. The four magnon term $N_k$ in Eq. (3) has the form

$$N_k = F_{k+1} - 2F_k + F_{k-1}\ . \quad (9)$$

This constitutes a closed system of equations for the determination of the dynamics of $n_k(t)$ for a given lattice temperature temporal profile $T_L(t)$.

The dynamics of the magnon population number $n_k(t)$ is obtained by solving Eq. (3) numerically with the parameters shown in Extended Data Table 1, the results of these numerical calculations are shown in Extended Data Fig. 1. The lattice temperature temporal profile $T_L(t)$ is assumed to be a step-like function, which changes from an elevated temperature $T_1$ to room temperature $T_0$ in the moment of time $t_{Off} = 0$ as is shown in the upper panel in Extended Data Fig. 1a. The simulated magnon densities $D_k n_k(t)$ are exemplarily plotted for four characteristic times in Extended Data Fig. 1b. The calculated magnon distributions $n_k(t)$, are fitted with the Bose-Einstein distribution function in order to obtain the corresponding chemical potential $\mu(t)$. The procedure of this fitting is as follows. First, we linearize the Bose-Einstein distribution Eq. (1):

$$\ln\left(\frac{1}{n(f,t)}+1\right) = \frac{hf}{k_B T_{\mathrm{eff}}(t)} - \frac{\mu(t)}{k_B T_{\mathrm{eff}}(t)}\ . \quad (10)$$

Consequently, the right-hand-side of Eq. (10) is a linear function of energy (frequency). We fit the low-energy area (below 50 GHz) of the magnon distribution. The fit converges at every time point, indicating that the low-energy magnons are in equilibrium at every instance of time. From fitting the simulated distributions at each moment of time, we obtain the time dependence of the chemical potential $\mu(t)$. The resulting dependence is shown in the middle panel in Extended Data Fig. 1a. In order to compare to the experimental data (see Fig. 2 in the main manuscript), we perform an integration of the magnon density $D_k n_k(t)$ over the same frequency range as in the experiment. The corresponding plot is shown in Extended Data Fig. 1a.

**Fabrication of the YIG/Pt nano-structures under investigation.** The investigated YIG/Pt strips are 500 nm (Fig.1, Fig. 2a-b and Fig. 3) and 1000 nm (Fig. 2c-e) wide and 5 μm long. The distance between the leads is 4 μm. The structures were fabricated using a YIG film of 70 nm thickness grown by means of Liquid Phase Epitaxy (LPE) on a (111) oriented Gadolinium Gallium Garnet (GGG) substrate[20]. Standard microwave-based ferromagnetic resonance (FMR) measurements yield a Gilbert damping parameter of $\alpha_{YIG} = 1.8\times10^{-4}$ for the bare out-of-plane magnetized YIG film with an inhomogeneous broadening of $\mu_0 H_0 = 0.1$ mT, and an effective saturation magnetization of $M_S = 123$ kA/m. Next, plasma-assisted cleaning was used to remove potential contaminations before the sample was transferred into a Molecular Beam Epitaxy (MBE) facility, where the sample was heated up to 200°C for 2 hours at a pressure of $p = 3.7\times10^{-9}$ mbar before a 7 nm thick Pt layer was deposited on top of the YIG film[33]. The deposited Pt layer was found to increase the Gilbert damping to $\alpha_{YIG/Pt} = 1.2 \times 10^{-3}$. The subsequent structuring of the YIG/Pt strips was achieved by Electron-Beam Lithography and Argon-Ion Milling[34] to a depth of approximately 40 nm. Afterwards the 150 nm thick gold contacts were structured by Electron-Beam Lithography and Electron-Beam evaporation, with a 10 nm thick Titanium seed layer. Then a focused Ion-Beam was used to remove the YIG adjacent to the structures and to polish the edges, resulting in a material removal to a depth of 300 nm. The resistance of the investigated devices is typically in the range of 600 Ohm for a width of 500 nm and in the range of 400 Ohm for a width of 1000 nm.

**Indicators of magnon condensate formation.** Formation of a spontaneous coherency caused by a thermodynamic phase transition in a bosonic system is considered[13,35] to be the most prominent property of a BEC of quasiparticles. It relates to the fact that "experimentally, with an energy distribution alone, it is difficult to tell whether particles really are in the same quantum state, or just close to it."[34] Unfortunately, in our experiment, the coherency of a magnon BEC cannot be proven directly due to the limited frequency resolution of the used BLS setup. Even the increased frequency resolution in a special BLS experiment in Ref. 12 was not high enough to prove the coherency directly. Nevertheless, using the micro-focused space-resolved BLS spectroscopy, the coherency of a magnon BEC has been proven by the demonstration of a stable standing wave pattern[36]. As well, the direct measurement of the coherency of a magnon BEC has been done for the SSE-mediated BEC using specialized microwave technique[17]. In addition, such phenomena as quantized vorticity[36] or Bogoliubov waves[37], being canonical features of both atomic and quasiparticle quantum condensates, can be recognized as indicators[38,39] of the magnon BEC formation. All these observations evidence the ability for the room-temperature magnon BEC in magnetic films[12,36,37] and nanowires[17]. At the same time, the proof of a BEC can be performed using its other general features. In our studies, we decided for the set of three indicators for the BEC, which directly followed from the Bose-Einstein distribution function given by Eq. (1). These are the following: (i) a threshold behavior, (ii) the spontaneous population of the lowest energy level, and (iii) the increase of the chemical potential up to the energy of this state. We show that all these three signs of condensate formation are present in our case. Clear threshold behaviours of BEC is shown in Fig. 3. Similar to Ref. 12, where the magnon BEC has been reported for the first time, here, a magnon accumulation at the lowest energy state within the frequency resolution of our setup is observed (Fig. 2a). Additionally, the extracted chemical potential does not only reach the minimum energy of the system (Fig. 2b) but fits perfectly to the presented theoretical predictions and a clear threshold behaviour is observed (Fig. 3).

**Measurements of the magnon density by time-resolved BLS spectroscopy.** In order to apply current pulses to the Pt nano-strip, a pulse generator (*Keysight 81160 A*) providing transition times down to 1 ns is connected to the sample using RF probes (*picoprobes*). For the BLS measurements[21], a laser beam with a wavelength of 457 nm and a power of 1.5 mW is focused onto the center of the structures. To be able to probe the ultra-thin YIG film covered by the reflective metal layer (Pt), the probing light is directed to the YIG film from the opposite sample side through the transparent Gallium Gadolinium Garnet (GGG) substrate. The laser-spot size is approximately 400 nm in diameter. The experiment was performed with a repetition rate of 1 µs. The intensity of the frequency-shifted and backscattered light is proportional to the magnon density[21]. The maximal magnon wave vector, which can be detected is given by the laser wavelength and the numerical aperture of the objective $N_A = 0.85$. In our setup, magnons with an in-plane wave vector up to 23.3 rad/μm can be detected. The frequency shifted light passes a three-pass tandem Fabry-Pérot interferometer and is detected frequency selectively by a single photon counting module with a resolution of $\Delta f_{\mathrm{BLS}} = 150$ MHz. So broad frequency band does not allow direct measurement of the BEC coherency[35] like it was done in Ref. 17 by means of electrical detection of the SSE-induced BEC but is very suitable for the observations of transient magnon dynamics reported here. The output of the optical detector is connected to a fast data acquisition module, which is synchronized with the applied pulses. The time resolution is limited by the time the scattered photons stay in the interferometer, which is determined by the finesse of the interferometer and the distance between the mirrors ($d = 5$ mm in the experiment), resulting in a time resolution of approximately $\Delta t = 2$ ns. An in-plane biasing field of $B_{\mathrm{ext}} = 188$ mT is applied, which is sufficiently large to magnetize the strips in the direction of the applied field. The orientation of the external field with respect to the strip is changed by turning the sample by 90 degrees.

**Measurements of magnon chemical potential by BLS spectroscopy.** The condition for the quasi-particle BEC is that the chemical potential is approaching the minimal energy of these quasi-particles. To subtract the chemical potential directly from our experiment, the following procedure was used. For simplicity, the chemical potential $\frac{\mu(t)}{h}$ in units of frequency is calculated from Eq. (1) in the approach $k_{\mathrm{B}}T \gg hf - \mu$ (corresponding to the Rayleigh-Jeans limit). Hence, the number of magnons $n_i$ occupying a certain mode of the frequency $f_i$ is given by $n_i = \frac{Dk_{\mathrm{B}}T}{hf_i - \mu}$. Since both, the first standing thickness mode and the fundamental mode obey the Bose-Einstein distribution, the chemical potential can be expressed as

$$\frac{\mu(t)}{h} = \frac{n_{\mathrm{1st\,mode}}(t) f_{\mathrm{1st\,mode}}(t) - n_{\mathrm{FM}}(t) f_{\mathrm{FM}}(t)}{n_{\mathrm{1st\,mode}}(t) - n_{\mathrm{FM}}(t)}. \quad (11)$$

Thereby $\frac{\mu(t)}{h}$ becomes a function of the number of magnons in both modes $n_{\mathrm{1st\,mode}}$, $n_{\mathrm{FM}}$ and of the corresponding frequencies $f_{\mathrm{1st\,mode}}$, $f_{\mathrm{FM}}$. For the case of thermal equilibrium, the chemical potential is zero and therefore, the number of magnons in one mode is given by $n_i = \frac{Dk_{\mathrm{B}}T}{hf_i} = b_i I_i$. Here, $I_i$ are the measured integrated BLS intensities and the factors $b_i$ express the BLS sensitivities for different modes. Thus, the factors $b_i$ can be calculated from the thermal BLS intensities ($t > 300$ ns in Fig. 2a, 2b) corresponding to a magnon system in equilibrium. Both, the measured BLS frequency of the fundamental mode $f_{\mathrm{FM}}$ and the chemical potential obtained experimentally with help of Eq. 11 are shown in Fig. 2b. In addition, the black line in Fig. 2b presents the time evolution of the BLS intensity integrated over the frequency range at the bottom of the magnon spectrum.

As can be seen from Fig. 2b, a good agreement with the simplified theoretical model in the middle panel of Fig. 1a is visible. The BLS intensity at the bottom of the magnon spectrum increases after the pulse. In addition, the accompanying increase of the chemical potential is verified experimentally. At its maximum, the chemical potential coincides within the error bars with the lowest magnon frequency. (Please note that lowest magnon frequency slightly shifts down during the current pulse due to the heating.) Hence, the criterion for the BEC – the equality of the chemical potential and the minimum energy of the system – is fulfilled.

Please note that an increase in the chemical potential is also observed during the pulse. This increase can be attributed to the spin Seebeck effect[17,24,25]. As it is reported by I. Barsukov and I. N. Krivorotov at a variety of international conferences, the spin Seebeck effect servs as the main mechanism of magnon BEC in their experiments[17]. However, in our experiments, the thermally induced spin injection from the Pt layer is not sufficient to reach the BEC condition. Even though the experiment was performed at various voltages and pulse durations, we observed the BEC always after the pulse, i.e., at times at which the gradient has already disappeared, and never during the pulse. The same experiment in the continuous regime also did not show any magnon generation in the lowest or any other observable magnon state. This can be related Therefore, in our experiment, the spin Seebeck effect can be seen just as a supportive mechanism for the BEC while the key mechanism is rapid cooling.

**Acknowledgements:**

This research has been supported by ERC Starting Grant 678309 MagnonCircuits, ERC Advanced Grant 694709 Super-Magnonics, by the DFG in the framework of the Research Unit TRR 173 “Spin+X” (Projects B01 and B07) and Project DU 1427/2-1, by the grants Nos. EFMA-1641989 and ECCS-1708982 from the National Science Foundation of the USA, and by the DARPA M3IC grant under the contract W911-17-C-0031.

**Author contributions:**

M.S. and D.B. performed the measurements and analyzed the experimental results. T.B. and A.V.C. supervised the measurements. T.B., P.P., A.A.S., A.V.C. planned the experiment. M.S., Bj.H., T.M. developed the experimental set-up. V.L. performed FMR characterizations and preliminary experiments. C.D. has grown the LPE YIG film. S.K. and E.Th.P. deposited the Pt overlayer. M.S., Bj.H., B.L., T.L., D.B. fabricated the structures under investigation. V.S.T. developed the quasi-analytical model of the magnon spectral redistribution. D.A.B., H.Yu.M.-S., V.S.T., A.N.S. performed the theoretical calculations. M.S. and F.H. performed the COMSOL simulations. Q.W. and P.P. performed the MuMax3 simulations. B.H. and A.V.C. led the project. All authors discussed the results and wrote the manuscript.

**Competing interests:** None declared.

**Affiliations:**

[1]Fachbereich Physik and Landesforschungszentrum OPTIMAS, Technische Universität Kaiserslautern, D-67663 Kaiserslautern, Germany.

[2]Department of Physics and Energy Science, University of Colorado at Colorado Springs, Colorado Springs, CO 80918, USA

[3]Graduate School Materials Science in Mainz, Staudingerweg 9, D-55128 Mainz, Germany.

[4]THATec Innovation GmbH, Bautzner Landstraße 400, D-01328 Dresden, Germany.

[5]Nano Structuring Center, Technische Universität Kaiserslautern, D-67663 Kaiserslautern, Germany.

[6]INNOVENT e.V. Technologieentwicklung, Prüssingstraße 27B, D-07745 Jena, Germany.

[7]Department of Physics, Oakland University, Rochester, MI 48309, USA.

[8]Faculty of Physics, University of Vienna, Boltzmanngasse 5, AT-1090 Wien, Austria.

[*]Corresponding author. Email: chumak@physik.uni-kl.de

# Supplementary Information

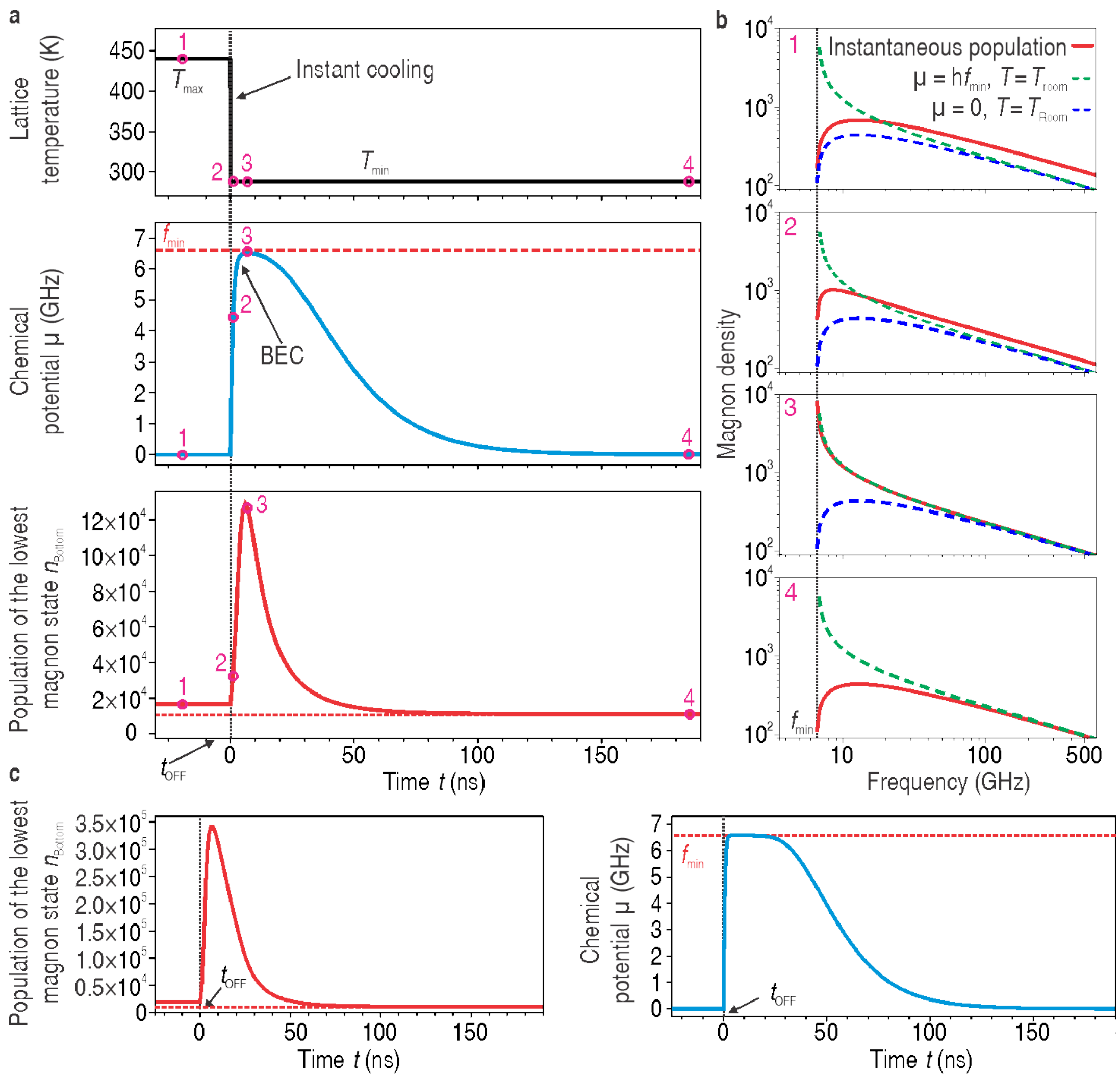

**Extended Data Fig. 1 | Detailed results of the numerical modelling of the BEC by rapid cooling. a,** Time evolution of the temperature of the phonon system (top panel), the magnon chemical potential (middle panel) and the magnon population at the lowest energy state (bottom panel). The moment in time of the instant cooling is marked by $t_{\mathrm{Off}}$ and the vertical dashed line. **b,** Frequency distributions of the quasi-particles density for different times. The red lines in panels 1–4 show the snapshots of the magnon densities (for the times marked by the pink dots in panel **a** calculated using the dynamics equations. Dashed blue lines show the steady-state room-temperature distribution calculated using Eq. (1) in the manuscript. The green dashed lines denote the distribution when the chemical potential μ is equal to the minimal magnon energy $hf_{\mathrm{min}}$. Panel 1: Stationary heated case before the instant cooling. Panel 2: Distribution just after the instant cooling. Panel 3: Snapshot of magnon density when the chemical potential μ reaches the minimal energy $hf_{\mathrm{min}}$, resulting in BEC. Panel 4: Final room temperature equilibrium state. **c,** Time evolution of the chemical potential and the magnon population at the lowest energy state calculated for a heating temperature of 510 K, which exceeds the threshold value of 440 K. The long time pinning of the chemical potential μ at $f_{\mathrm{min}}$ is clearly visible for the case of the stronger heating.

| Parameter | Value | |
|---|---|---|
| Width of the strip | 500 nm | 1000 nm |
| Current pulse duration $\tau_p$ | 120 ns | 300 ns |
| Applied voltage $U$ | 0.9 V | 1.05 V |
| Resistance $R$ | 600 Ohm | 403 Ohm |
| Room temperature $T_0$ | 288 K | 288 K |
| Elevated temperature $T_1$ | 440 K | 490 K |
| Gilbert damping constant $\alpha_G$ | $1.5\times10^{-3}$ | $1.5\times10^{-3}$ |
| Maximal frequency $f_{max}$ | 600 GHz | 600 GHz |
| Frequency step $\Delta f$ | 250 MHz | 250 MHz |
| Four-magnon scattering efficiency $C$ | 0.7 | 0.7 |

**Extended Data Table 1 | Parameters used for the calculations of the magnon density**. The table shows the parameters according to the developed quasi-analytical theoretical model for two different experimentally investigated strips.

**Spin-wave dispersion in the YIG nano-structures.** The dispersion relations were obtained by means of micromagnetic simulations, which were performed using MuMax3[1] – see Extended Data Fig. 2. The following parameters were used for the micromagnetic simulations. The size of the waveguide is 20 µm × 500 nm × 70 nm (length × width × thickness). The mesh was set to 10 nm × 10 nm × 70 nm. The following parameters for YIG were used: Saturation magnetization $M_s$ = 123 kA/m, exchange constant $A$ = 8.5 pJ/m, and Gilbert damping $\alpha = 2 \times 10^{-4}$. In the simulations, the damping at the ends of the waveguides was set to increase exponentially to 0.5 in order to eliminate spin-wave reflections. The external field is 188 mT for both, the Backward Volume and the Damon-Eshbach geometries[18]. In order to excite spin-wave dynamics in the waveguide, a sinc-function-shaped field pulse was applied to a 50 nm wide area in the center of the waveguide. To gain access to both, odd and even spin-wave-width modes, this area was slightly shifted away from the center along the short axis of the waveguide. The sinc field is given by $\mathbf{b}_z = b_0 \,\mathrm{sinc}\,(2\pi f_c t)$, with an oscillation field $b_0$ = 1 mT and a cutoff frequency $f_c$ = 20 GHz. The out-of-plane component $M_z(x, y, t)$ of each cell was collected over a period of $t$ = 50 ns and stored in $t_s$ = 12.5 ps intervals, which results in a frequency resolution of $\Delta f = 1/t = 0.02$ GHz, whereas the highest resolvable frequency was $f_{max} = 1/(2t_s) = 40$ GHz. The fluctuations in $m_z(x, y, t)$ were calculated for all cells via $m_z(x, y, t) = M_z(x, y, t) - M_z(x, y, t = 0)$, where $M_z(x, y, t = 0)$ corresponds to the ground state. To obtain the spin-wave dispersion curves, a two-dimensional fast Fourier transformation in space and time has been performed[2].

[1] Vansteenkiste, A., et al. The design and verification of MuMax3, *AIP Adv.* **4**, 107133 (2014).

To visualize the dispersion curve, Extended Data Fig. 2 shows a 3D grey-scale map of the spin-wave intensity, which is proportional to $m_z^2\,(k_x, f)$, in logarithmic scale as a function of $f$ and $k_x$. In both configurations, several width modes are visible, of which the first three are marked in the graphs in Extended Data Fig. 2. Since the first three width modes are at similar frequencies, they cannot be distinguished in the experiment. The energy minimum, where the generation of the Bose Einstein condensate is expected, is in both cases within the wave vector range accessible with Brillouin-Light Scattering (BLS) spectroscopy ($k_x = 13$ rad/µm for **B**||**y** and $k_x = 0$ rad/µm for **B**||**x**). The frequency of the minimum of the magnon spectrum coincides with the ferromagnetic resonance frequency ($k = 0$) in the case of the Damon-Eshbach **B**||**y** geometry and differs by a value of 380 MHz in the case of the Backward Volume geometry **B**||**x**. This difference is on the order of magnitude of the frequency resolution of the BLS spectroscopy. In general, the calculated frequency at the bottom of the magnon spectrum agrees well with the frequency of the BEC peak observed in the experiment taking into account the decrease of the saturation magnetization due to the increase in temperature within the YIG nano-structure.

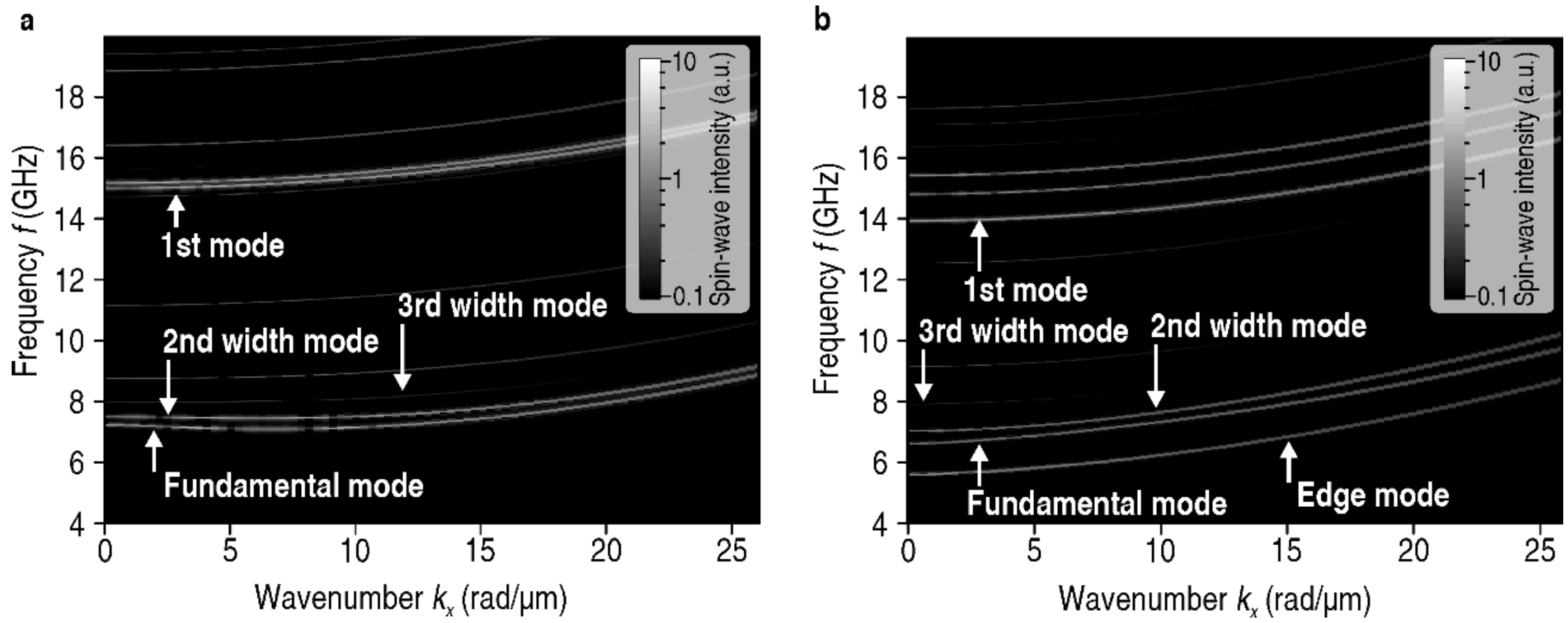

**Extended Data Fig. 2 | Simulated magnon dispersions. a**, Simulated spin-wave intensity shown in a logarithmic black-to-white scale as a function of the frequency $f$ and the wavenumber $k_x$ along the waveguide for **B**||**x. b,** Simulated spin-wave intensity as a function of the frequency $f$ and the wavenumber $k_x$ along the waveguide for **B**||**y**.

[2] Venkat G., et al. Proposal of a standard micromagnetic problem: Spin wave dispersion in a magnonic waveguide, *IEEE Trans. Magn.* **49**, 524 – 529 (2013).

**Simulation of time evolution of YIG/Pt structure by COMSOL Multiphysics.** The numerically calculated time evolution of the temperature of the YIG/Pt strips was determined by solving a 3D heat-transfer model of the experimental set-up with the COMSOL Multiphysics Software using the heat transfer module and the electric currents module.

Hereby, the conventional heat conduction differential equation and the differential equations for current conservation are solved taking into consideration the boundary conditions applied to the model as well as the material parameters of the materials used. The model comprises a 9 μm × 4.5μm × 6.5 μm large volume which includes half of the strip, exploiting the symmetry of the system with respect to its long axis. The simulated geometry includes the YIG/Pt strip, a part of the Au leads, which have a width of 1000 nm where they meet the strip and a width of 8 μm at a distance of 1 μm to the strip. In addition, the removal of the material by focused Ion-Beam with a width of 4 μm at the sides of the strip with a depth of 300 nm is taken into account. The used material parameters can be found in Table Extended Data Table 2. The electrical conductivity and the corresponding temperature coefficient of Pt were measured experimentally, whereas the other parameters are taken from the referenced literature or from the COMSOL library. For the heat-transfer model, the boundary on the edges were set to the symmetry boundary condition for the according edges and to open boundary for the others. For the electric-currents model they were set to electric insulation, except for the two faces of the leads where the electric potential is applied. The Joule heating is modelled for all metallic layers in the system. The simulation differs from the experiment by the fact that the convective cooling due to the surrounding air as well as the laser heating are not implemented.

Extended Data Fig. 3 shows the obtained temperature profile in the center of the Pt overlayer. The profile was obtained for a 300 ns long pulse with an amplitude of $U_{\mathrm{Sim}} = 1.872$ V (corresponding to a set voltage of $U = 1.05$ V at the pulse generator for a 50 Ohms load resistance in the experiment) and transition times of 1 ns at the edges. It can be seen that the temperature increases rapidly just after the current pulse is applied to Pt but tends to saturate. After a time of approximately 300 ns the temperature still grows but the heating has already strongly decelerated allowing for the formation of the quasi-equilibrium between the magnon and phonon temperatures[3] considered in the theoretical model. The maximal temperature reached at the end of the applied current pulse in this particular case is 491 K.

[3] Agrawal, M., et al. Direct measurement of magnon temperature: New insight into magnon-phonon coupling in magnetic insulators. *Phys. Rev. Lett.* **111**, 107204 (2013).

| **Parameter** | **Material** | **Value/ Source** |
|---|---|---|
| Density | YIG | 5170 kg $m^{-3}$<br>Clark, A. E. & Strakna, R. E. Elastic constants of single-crystal YIG, *J. Appl. Phys.* **32**, 1172 (1961) |
| Heat conductivity | YIG | 6 W $m^{-1}K^{-1}$<br>Hofmeister, A. M. Thermal diffusivity of garnets at high temperature, *Phys. Chem. Minerals* **33**, 45-62 (2006) |
| Heat capacity | YIG | 570 J $kg^{-1}$ $K^{-1}$<br>Guillot, M., Tchéou, F., Marchand, A., Feldmann, P. & Lagnier, R., Specific heat in Erbium and Yttrium Iron garnet crystals, *Z. Phys. B – Condensed Matter* **44**, 53-57 (1981) |
| Density | GGG | 7080 kg $m^{-3}$<br>Hofmeister A. M. Thermal diffusivity of garnets at high temperature, *Phys. Chem. Minerals* **33**, 45-62 (2006) |
| Heat conductivity | GGG | 7080 kg $m^{-3}$<br>Hofmeister A. M. Thermal diffusivity of garnets at high temperature, *Phys. Chem. Minerals* **33**, 45-62 (2006) |
| Heat capacity | GGG | 7080 kg $m^{-3}$<br>Hofmeister A. M. Thermal diffusivity of garnets at high temperature, *Phys. Chem. Minerals* **33**, 45-62 (2006) |
| Density | Pt | 21450 kg $m^{-3}$<br>Lide, D. CRC Handbook of Chemistry and Physics, 89th ed. (Taylor & Francis, London, 2008) |
| Electrical conductivity | Pt | $1.41 \times 10^{6}$ S $m^{-1}$ (Extended Data Fig. 3), $1.90 \times 10^{6}$ S $m^{-1}$ (Extended Data Fig. 4), $1.79 \times 10^{6}$ S $m^{-1}$ (Extended Data Fig. 5a)<br>Measured |
| Resistance-Temperature Coefficient | Pt | $7.135 \times 10^{-4}$ $K^{-1}$<br>Measured |
| Thermal Conductivity | Pt | 22 W $m^{-1}$ $K^{-1}$<br>Yoneoka, S., et al. Electrical and thermal conduction in atomic layer deposition nanobridges down to 7 nm thickness, *Nano Lett.* **12**, 683-686 (2012) |
| Heat capacity | Pt | 130 J $kg^{-1}$ $K^{-1}$<br>Furukawa, G. T., Reilly, M. L., Gallagher, J. S. Critical Analysis of Heat Capacity data and evaluation of thermodynamic properties of Ruthenium, Rhodium, Palladium, Iridium, and Platinum from 0 to 300K. A survey of the literature data on Osmium, *J. Phys. Chem. Ref. Data* **3**, 163 (1974) |
| Density | Au | 1900 kg $m^{-3}$<br>COMSOL Library |

| Electrical conductivity | Au | $4.1 \times 10^{7}$ S $m^{-1}$<br>COMSOL Library |
|---|---|---|
| Thermal Conductivity | Au | 190 W $m^{-1}$ $K^{-1}$<br>Langer, G., Hartmann, J. & Reichling M. Thermal conductivity of thin metallic films measured by photothermal profile analysis**,** *Rev. Sci. Instrum.* **68***,* 3 (1997) |
| Heat capacity | Au | 130 J $kg^{-1}$ $K^{-1}$<br>Geballe, T. H., & Giauque, W. F. The heat capacity and entropy of Gold from 15 to 300°K, *J. Am. Chem. Soc.* **1952***,* 74) |
| Density | Ti | 4500 kg $m^{-3}$<br>COMSOL Library |
| Electrical conductivity | Ti | $2 \times 10^{6}$ S $m^{-1}$<br>COMSOL Library |
| Thermal Conductivity | Ti | 22 W $m^{-1}$ $K^{-1}$<br>Ho, C. Y., Powell, R. W. & Liley, P. E. Thermal conductivity of the elements, J. Phys. Chem. Ref. Data **1**, 2 (1972) |
| Heat capacity | Ti | 521.4 J $kg^{-1}$ $K^{-1}$<br>Kothen, C. W. & Johnston, H. L. Low Temperature Heat Capacities of Inorganic Solids. XVII. Heat Capacity of Titanium from 15 to 305°K. *J. Am. Chem. Soc.* **75**, 3101 (1953) |
| Heat capacity | Air | $1047.6366 - 0.3726\,T + 9.4530 \times 10^{-4}\,T^{2} - 6.0241 \times 10^{-7}\,T^{3} + 1.2859 \times 10^{-10}\,T^{4}$ J $kg^{-1}$ $K^{-1}$<br>COMSOL Library |
| Thermal Conductivity | Air | $-0.0023 + 1.1548 \times 10^{-4}\,T - 7.9025 \times 10^{-8}\,T^{2} + 4.1170 \times 10^{-11}\,T^{3} - 7.4386 \times 10^{-15}\,T^{4}$ W $m^{-1}$ $K^{-1}$<br>COMSOL Library |
| Density | Air | $8.53\,T^{-1}$ kg $m^{-3}$<br>COMSOL Library |

**Extended Data Table 2 | Parameters used for the COMSOL Simulations.**

As it is seen in Extended Data Fig. 3a, switching off the current pulse results in a fast decrease in temperature of the YIG/Pt nano-structure due to the thermal diffusion in the quasi-bulk surroundings. The cooling rate is not constant and decreases with time. Moreover, our simulations further revealed (not shown) that the cooling rate depends on the duration of the applied current pulse which is associated with the increase in temperature of the surroundings of the nano-structure for long current pulses.

To prove the crucial role of the fast cooling for the BEC, additional measurements were performed where the current pulses were switched off with fall times of 50 ns and 100 ns – see Fig. 2 in the main text of the manuscript. It is evident that the condensate disappears for long fall times, i.e. slow cooling rates. For reference, Extended Data Fig. 3 shows the corresponding simulated temperature evolutions in the phonon system. The maximal cooling rates indicated in Extended Data Fig. 3b clearly show that a cooling rate of 2 K/ns is already too slow to trigger the magnon BEC, while a rate of 20.5 K/ns is still fast enough.

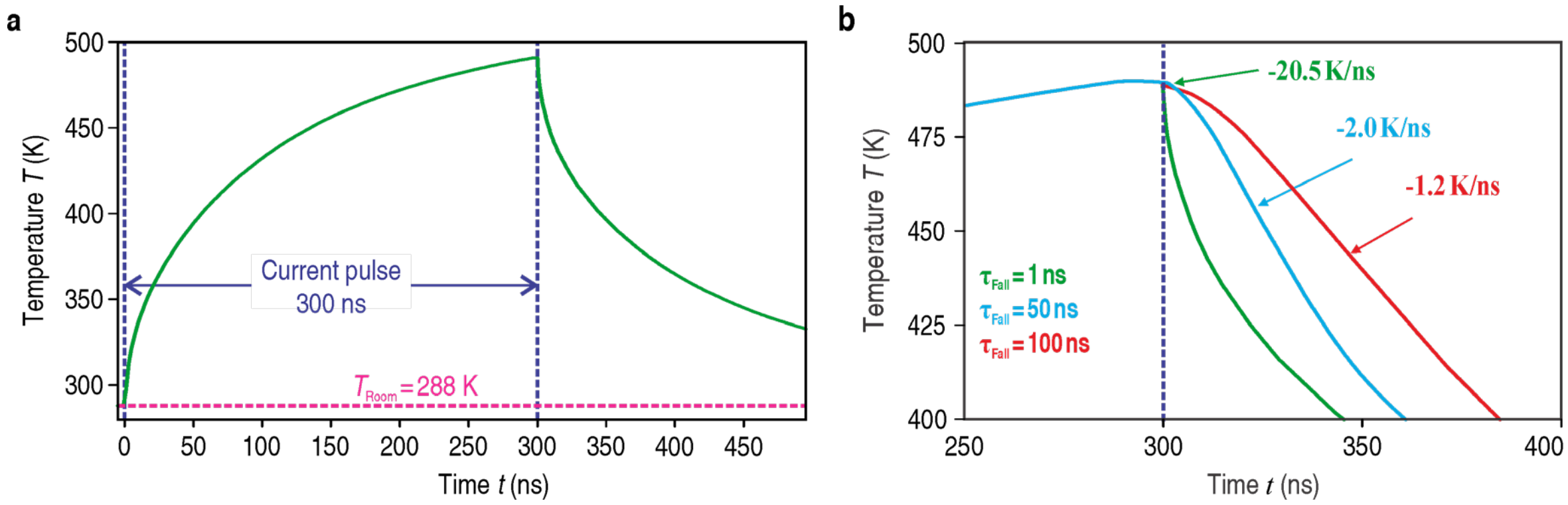

**Extended Data Fig. 3 | Results of the COMSOL simulations for different fall times. a,** Simulated temperature as a function of time in the middle of the Pt overlayer for a 300 ns long pulse with a voltage of 1.05 V applied to a 1 μm wide waveguide. The fall time of the pulse was 1 ns. The dashed blue lines indicate the time when the pulse is present. The room temperature is marked with the dashed pink line. **b,** Simulated temperature as a function of time for different fall times $\tau_{Fall} = 1$ ns (green line), $\tau_{Fall} = 50$ ns (blue line), $\tau_{Fall} = 100$ ns (red line). The maximal time derivatives are marked at their respective time of occurrence.

To support the COMSOL simulations an additional experiment is performed, measuring the temperature of the Platinum layer time-resolved by means of electrical measurements. First, $\tau_P = 300$ ns long DC-pulses with an amplitude of $U = 0.85$ V are applied to a 1 μm wide YIG/Pt waveguide. The corresponding BLS-spectrum is shown in Extended Data Fig. 4b. The frequency shift during the pulse indicates the heating of the magnetic system, whereas the magnon BEC is observed after the pulse is switched off. To get access to the time-resolved resistivity (which depends on the temperature) during the applied pulse instead of the DC-pulse, a RF-pulse with the same duration and a frequency of 12.4 GHz is applied to heat up the sample. The RF-power was adjusted in such a way, that Joule heating results in the same steady state temperature as for a continuous applied voltage of $U = 0.85$ V. Thus, the resistance of the Pt layer was found to increase from 319.5 Ohm to 372.1 Ohm when either a continuous DC-voltage of 0.85V or a continuous RF signal with a power of 18.84 dBm and a frequency of 12.4 GHz is applied. Hence, the RF-power of 18.84 dBm results in the same Joule heating of the Platinum layer as a DC-voltage of 0.85 V. Then, a continuous DC-voltage of 30 mV, which is used to measure the resistance change and a 300 ns long RF-pulse are applied in parallel using a DC-block and a RF-circulator. From the changed voltage drop on the sample, which is measured with an oscilloscope, the temperature is calculated. The resulting temperature profile is shown in Extended Data Fig. 4a (black curve). In addition, we have performed the same COMSOL simulation as for all other shown simulated temperature profiles. Only the corresponding conductivity of the Pt-layer was matched to the measured resistivity of the specific structure. The results are as well depicted in Extended Data Fig. 4a (blue curve). The perfect agreement with the measured temperature of the Platinum layer shows that the used COMSOL model can precisely describe the temperature dynamics of the investigated structures.

Further, from the BLS measurement the temperature is derived by considering the frequency shift determined by BLS and the Kittel equation. The resulting temperature profile is shown in Extended Data Fig. 4a (red points) and deviates from the simulated and electrically measured temperature during the pulse and directly after the pulse. At times the BEC already disappeared the temperature profiles coincide. This difference can be since the BLS-frequency during the pulse is not only given by the number of magnons, but also by the flattening of the dispersion, due to a temperature induced change of the magnetic parameters. Further, it might be influenced by a strong thermal gradient, which potentially changes the mode profiles and frequencies. At the time the BEC occurs, a too high temperature is measured by BLS, which indicates ones more, that magnons condense to the lowest energy state.

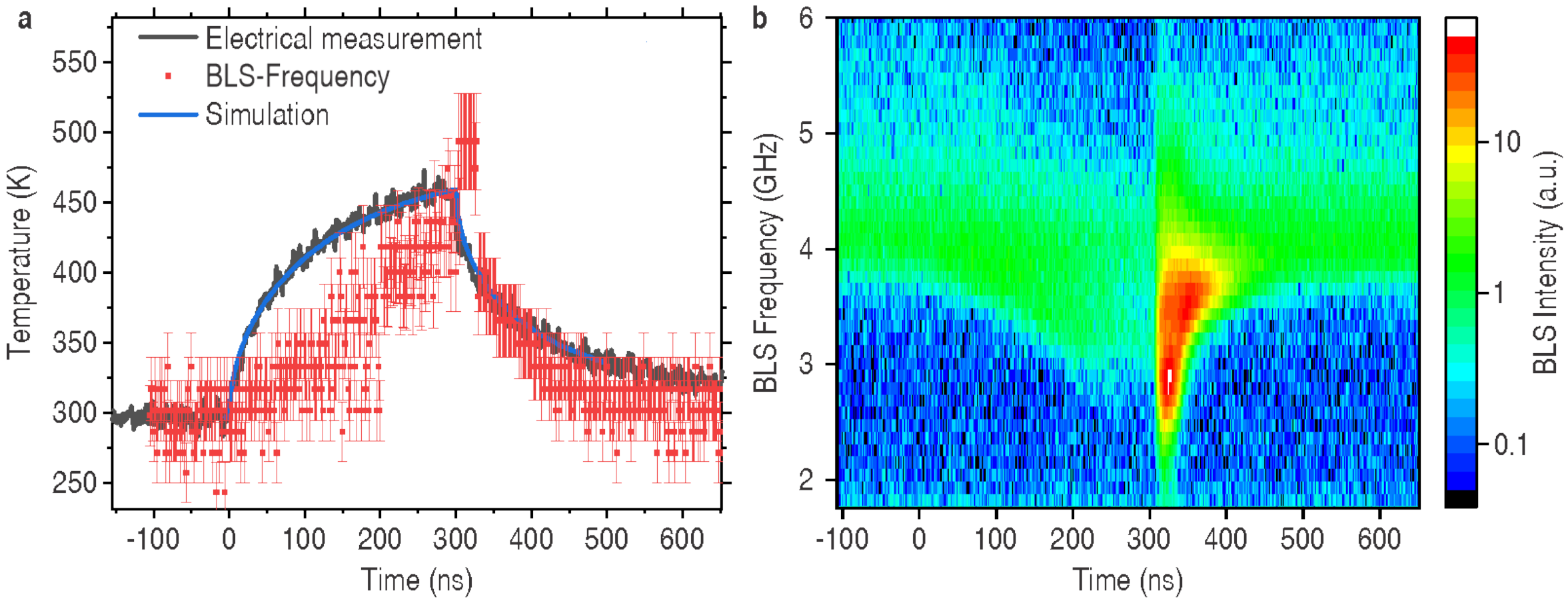

**Extended Data Fig. 4 | Temperature profiles derived by electrical measurements, COMSOL and BLS | a,** Temperature as a function of time measured electrically (black curve) by means of resistivity change for an applied RF-pulse with a frequency of $f = 12.4$ GHz, a power of $P_{RF} = 18.84$ dBm and a duration of $\tau_P = 300$ ns, obtained by COMSOL simulations (blue curve) for an applied voltage of $U = 0.85$ V and a pulse duration of $\tau_P = 300$ ns, derived from the shift of the BLS-frequency (red data points) **b,** Time resolved BLS spectrum for the temperature profiles shown in **a**, for the case when a pulse of an amplitude of $U = 0.85$ V and a duration of $\tau_P = 300$ ns is applied.

**Thermal dynamics and the spin Seebeck effect.** Another possible contribution to the observed BEC can be the spin-orbit-mediated spin Seebeck effect as it was shown in Ref. 17. To clarify the role of the SSE, we extracted the temperature profile across the YIG/Pt structure and simulated the temporal evolution of the temperature and temperature gradient in the YIG film – see Extended

Data Fig. 5. The simulation was performed for the experiment shown in Fig. 2b in the main manuscript. The current pulse duration was set to $\tau_P = 120$ ns. The pulse voltage was of $U_{Sim} = 1.662$ V, which corresponded to a set voltage of $U = 0.9$ V at the pulse generator for a 50 Ohms load resistance in the experiment. A maximum temperature of $T = 443$ K was reached at the end of the pulse. One can see that the gradient can reach values up to $3.5 \times 10^8$ K/m. However, the temporal evolution of the gradient shows that it is formed within few nanoseconds after the current pulse is applied, stays practically constant during the pulse and, finally, disappears within few nanoseconds after the pulse is switched off. Thus, at the moment of BEC in our experiment, there is no thermal gradient present as well as a possible contribution from the SSE. At the same time, we can assume that the SSE contributes to the increase in the magnon chemical potential during the current pulse is applied[17]. However, in course of time, such a contribution should decrease with increase in the YIG temperature due to a strong rapid temperature-dependent reduction in the SSE magnitude (see Fig. 3c in Ref. 27).

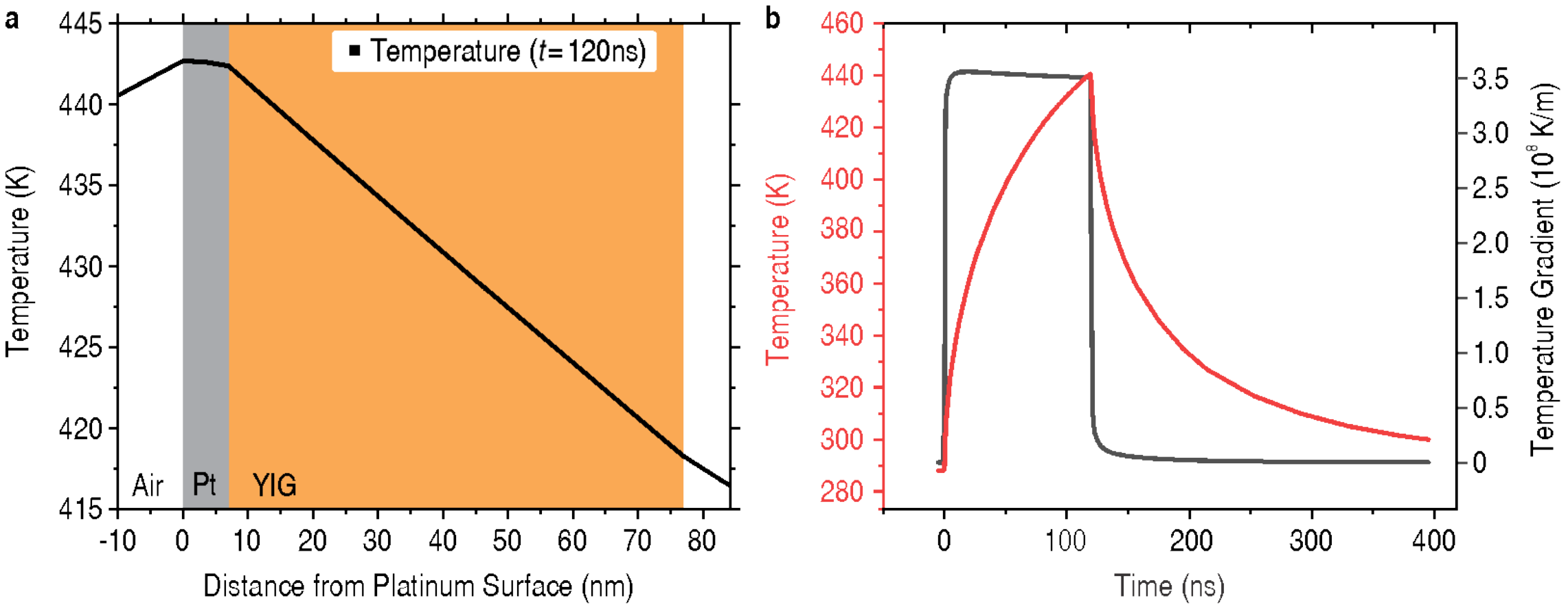

**Extended Data Fig. 5 | Temperature profile and time evolution of temperature gradient.** Simulated temperature (red curve, left axis) and temperature gradient (black curve, right axis) as a function of time. The profiles were simulated with COMSOL for the parameters corresponding to the experiment shown in Fig. 2b in the main manuscript.

**Dependence of the BEC on the applied magnetic field.** The described magnon BEC is driven by the change of the temperature of the phonon system. In order to prove this, a set of measurements using different structure sizes of width of 500 nm and 1000 nm, of different scanning laser powers in the range from 0.75 mW to 3.3 mW, and different scanning points on the structure were performed. The experimental finding of the BEC, which manifests itself as a strong magnon peak at the bottom of the spectrum, could always be confirmed. In particular, the threshold for BEC is found to be independent of the applied external field. Measurements of the threshold for a pulse duration of $\tau_P = 18$ ns are performed in the range from 110 mT to 270 mT. The maximal integrated BLS intensities of the fundamental mode can be seen in Extended Data Fig. 6.

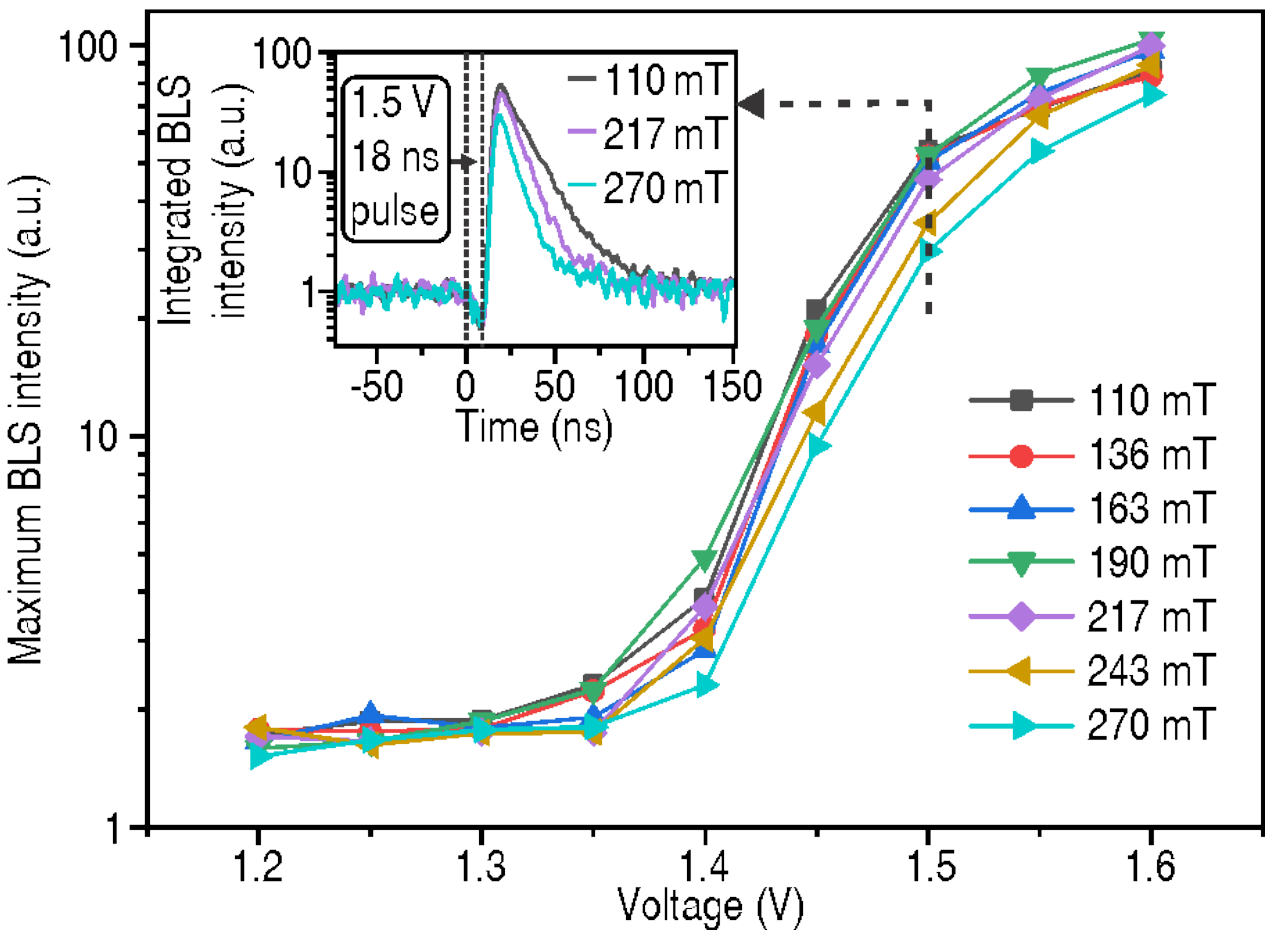

**Extended Data Fig. 6 | Dependency of the threshold on the external field.** Threshold curves for different external fields in the range from 110 mT to 270 mT. The field is applied in-plane along the short axis of a 500 nm wide strip. The inset shows exemplary time traces for a supercritical voltage of $U = 1.5$V and for three different fields. The threshold is found to be approximately constant over the measured field range, whereas the magnon lifetime decreases for higher values of the applied magnetic field.

In addition, the inset exemplary shows time traces for different field values and an applied voltage of $U$ = 1.5 V. From the threshold curves, it can be seen that the threshold of the BEC does not depend on the external field, whereas the inset shows that the magnon lifetime is decreased for higher fields, as expected. The measured lifetime is 21 ns and is small compared to the lifetime of magnons in YIG films of µm thicknesses, used in previous experiments[12,14,15,36,37]. It is known, that for YIG films with thicknesses in the nm-range, the spin-wave damping is larger, and consequently, the lifetime is smaller. Additionally, due to the spin pumping effect to the Pt layer and due to distortions induced in YIG sample by the structuring process the effective magnetic damping is enhanced.

Furthermore, the fact that there is no dependence of the phenomenon on the direction of the external field was confirmed by changing the direction of the applied field. Extended Data Fig. 7 shows the corresponding time resolved BLS spectra in the case that the external field is applied perpendicular (Extended Data Fig. 7a, **B**||**y**) or parallel (Extended Data Fig. 7c, **B**||x) to the long axis of the strip for similar pulse durations of $\tau_P = 120$ ns and $\tau_P = 150$ ns. Further, the integrated intensity of the fundamental mode is shown for both cases in Extended Data Fig. 7b (**B**||**y**) and 7d (**B**||**x**).

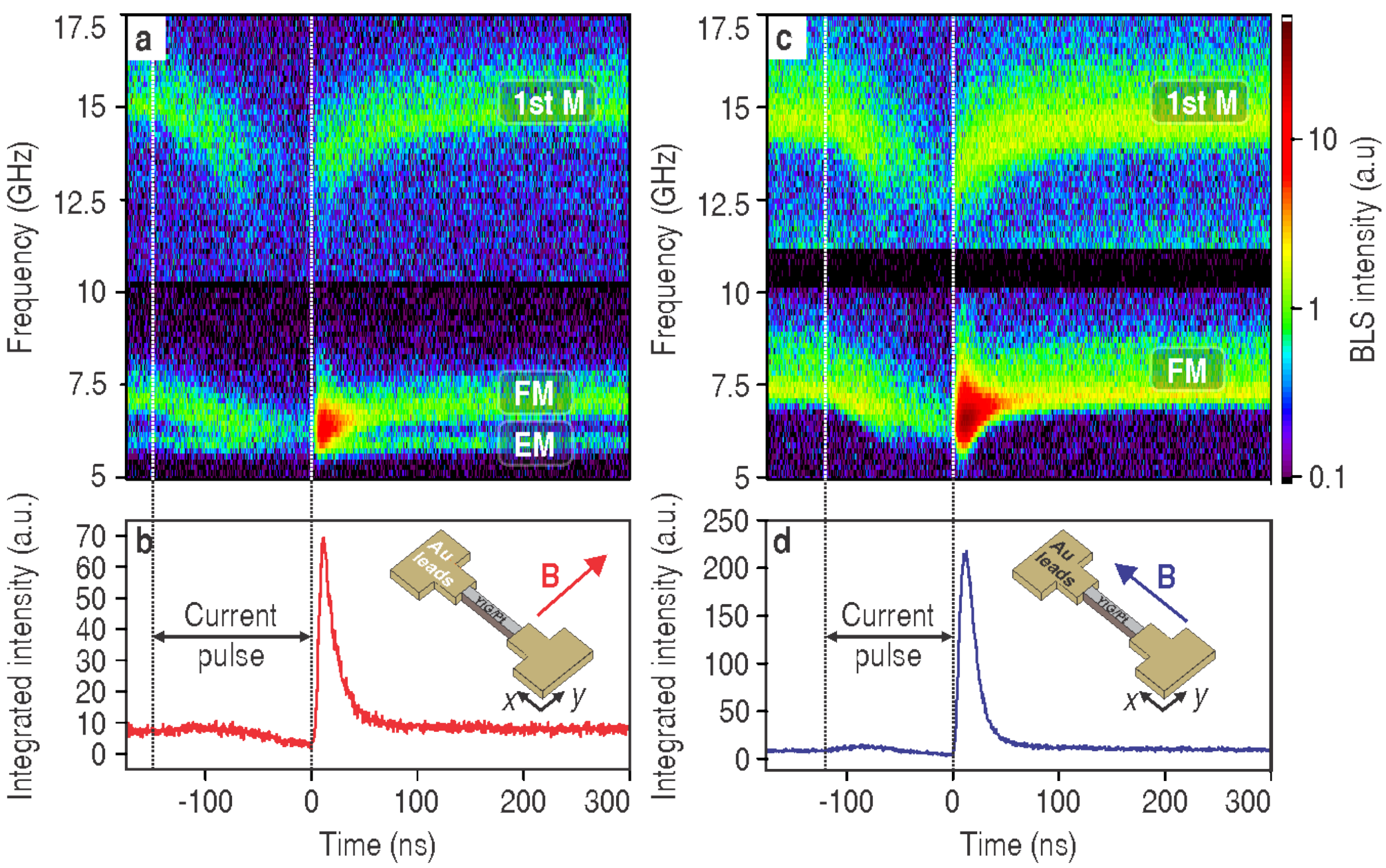

**Extended Data Fig. 7 | BEC by rapid cooling for different geometries of the external field.** **a,** BLS spectrum as a function of time. The BLS signal (color-coded, log scale) is proportional to the density of magnons. FM indicates the fundamental mode, EM – the edge mode, and 1st M – the first thickness mode. The vertical dashed lines indicate the start and the end of the pulse ($\tau_P = 150$ ns, $U = 0.9$ V). The external field was parallel to the short axis of the strip (**B**||**y**). **b**, BLS spectrum as a function of time for the case when the external magnetic field is parallel to short axis of the strip (**B**||**x**), ($\tau_P = 120$ ns, $U = 0.9$ V). **c, d,** Normalized magnon intensity integrated from 4.95 GHz to 8.1 GHz as a function of time for the cases in **a** and **b**. The insets show the sample and measurement geometry.

The frequencies of the fundamental modes are slightly larger if the magnetic field is applied parallel to the long axis of the strip. This is associated with the demagnetization fields that result in a smaller internal field when the strip is magnetized along its short axis. In addition, the inhomogeneity of the demagnetization fields leads to the formation of an edge mode with a frequency below the fundamental mode of the strip. The edge mode is visible in the experiment (see. Fig. Extended Data Fig. 7a) as well as in the simulations (see Extended Data Fig. 2b). The magnon BEC is observed in both cases at the bottom of the band of the waveguide modes, i.e., the lowest frequency of the fundamental mode. Even though the edge mode is the total energy minimum at room temperature equilibrium, it cannot be separated in the experiment from the fundamental mode during the BEC since, in contrast to the other modes, its frequency is not significantly decreased during the pulse. The reason for this is that the frequency of the edge mode strongly depends on the strength of the demagnetization fields. These fields are reduced as well when the saturation magnetization is decreased due to Joule heating. This behaviour was confirmed by additional MuMax3 simulations (not shown). The BEC for both field directions excludes a contribution from the Spin Hall Effect to the observed condensation, since it would only lead to an efficient injection of spin current when the magnetization is perpendicular to the direction of the electric current[26,28,29]. Moreover, since the phenomenon does neither depend on the direction nor on the value of the applied magnetic field (and thus also not on the frequency of the BEC), we can conclude that the Oersted fields generated by the eclectic current in the Pt overlayer do not play any sizable role in the experiments.

**Dependence of the BEC threshold on the current pulse duration.** The threshold voltages are determined experimentally for different pulse durations $18\,\text{ns} \leq \tau_P \leq 175\,\text{ns}$. For each pulse duration, BLS measurements were performed for different voltages. The maximum of the BLS intensity integrated over the frequency range of the fundamental mode was extracted and plotted as a function the applied voltage for each pulse duration. This yields graphs similar to the depiction in Fig. 3b in the main manuscript. The corresponding threshold voltage is determined by fitting the slope linearly and calculating the intercept with the averaged subcritical (thermal) BLS intensity. The extracted threshold voltages are shown in Extended Data Fig. 8.

As can be seen, the threshold voltage increases for the shortest pulse durations, whereas it saturates for the longest ones. This is due to the finite timescale of the Joule heating, which is in agreement with our model. The threshold temperature (i.e., the critical magnon density) should be independent of the applied voltage. As expected, the COMSOL simulations yielded similar values for the threshold temperature for all $\tau_P > 70$ns (blue points in Extended Data Fig. 8).

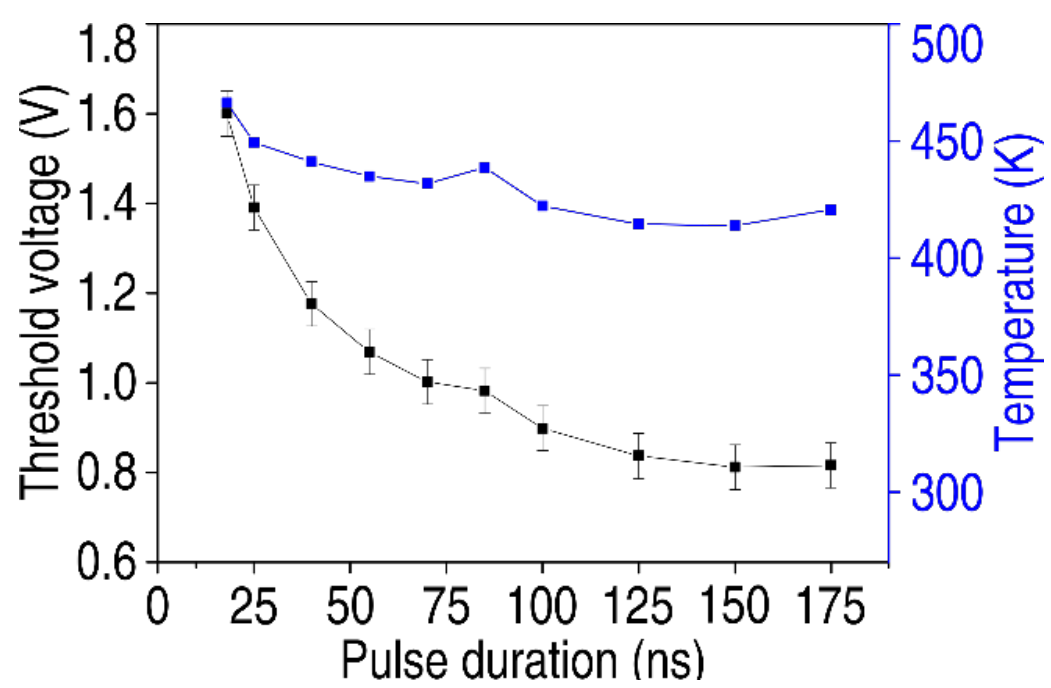

**Extended Data Fig. 8 | Dependency of the threshold on the pulse duration.** Threshold voltage as a function of the pulse duration (black squares) and corresponding simulated temperatures at the end of the pulses (blue squares).

The slight increase in the threshold temperature for the shorter pulses is likely related to incomplete thermalization of low energy magnons for such short times. Note, that in our experiment the spin-lattice relaxation time for the low energy magnons is about $21 \pm 4$ ns. The solid line in Extended Data Fig. 9 shows the calculated time evolution of the chemical potential for the current pulse of 20 ns, which heats the sample to the maximal temperature of 440 K. In such a case, the chemical potential induced by the rapid cooling might not reach the bottom of the magnon spectrum $hf_{\text{min}}$ and no BEC occurs.

Nevertheless, as it is supported by the experimental findings shown in Extended Data Fig. 8, the incomplete equilibration of magnon and phonon systems before the beginning of the fast cooling process does not impede the BEC phenomenon. The critical magnon density can still be achieved at a higher phonon temperature by application of a stronger heating current. The dashed line in Extended Data Fig. 9 shows that the BEC takes place for the 20 ns long pulse if the maximal sample temperature is increased to 490 K.

It is worth mentioning, that the "rapid heating" of the magnons' environment just after the application of the current pulse should act as an inverse of the rapid cooling process and, thus, lead to the decrease in the chemical potential on a time interval comparable with the magnon thermalization time. This effect is clearly visible in the Extended Data Fig. 9 but is not observed in our experiments. The reason could be the following. In general, the spin Seebeck effect leads to an increase of magnon chemical potential and the magnon dynamics **during** the current pulse is determined by the interplay of two counter-acting processes: "rapid heating" and SSE. Proper theoretical description of the magnon dynamics in this interval would require development of a completely new theory that would treat these effects on equal footing and is outside the scope of the present work. However, simple estimation of the SSE-induced effects just after the current pulse is switched on shows, that temperature gradient might lead to magnon chemical potential $\mu \sim 5.25$ GHz and should rapidly decrease afterwards due to the increase in the YIG temperature[27]. Thus, it looks reasonable that the "rapid heating" decrease in the chemical potential is compensated or even overcompensated by the pulsed-like contribution of the spin Seebeck effect.

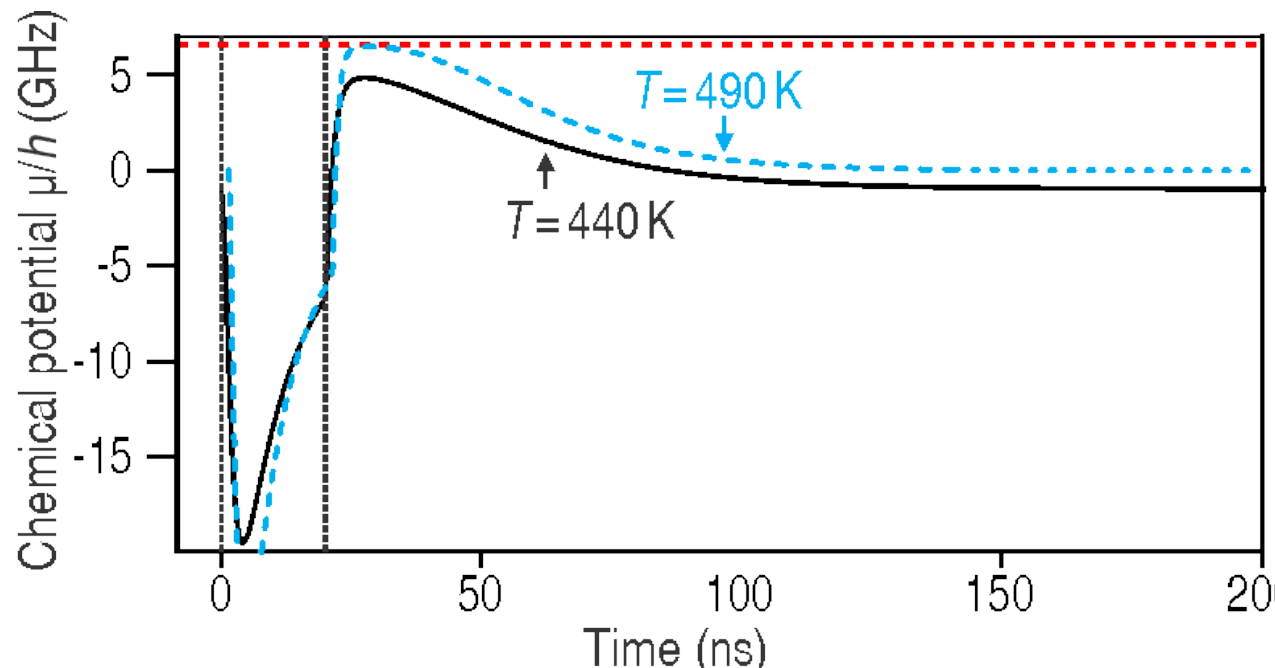

**Extended Data Fig. 9 | Time evolution of the chemical potential for a short heating pulse.** The pulse duration $\tau_P = 20$ ns. The initial decrease in the chemical potential is caused by the rapid heating of the sample. The incomplete thermalization of the magnon sub-system ($\mu < 0$) at the end of the pulse increases the threshold temperature of the BEC formation from 440 K to 490 K.

**Dependence of the BEC threshold on the fall time of the current pulses.** The influence of the cooling rate on the magnon BEC is investigated by applying $\tau_P = 50$ ns long current pulses at a 1 µm wide YIG/Pt structure. The fall times of the applied pulses varied in the range of $1 \text{ ns} \leq \tau_{Fall} \leq 100 \text{ ns}$. Extended Data Figure 10 shows the integrated BLS intensity as a function of the fall time. The strong increase for shorter fall times is clearly visible (above-threshold regime), whereas for fall times larger than the spin wave lifetime the accumulation at the bottom of the spectrum vanishes indicating that the BEC threshold is not reached. This supports the conclusion from the main manuscript, that a rapid cooling of the phonon system is crucial to achieve the magnon BEC and the required cooling rate is mainly determined by the lifetime of the low energy magnons.

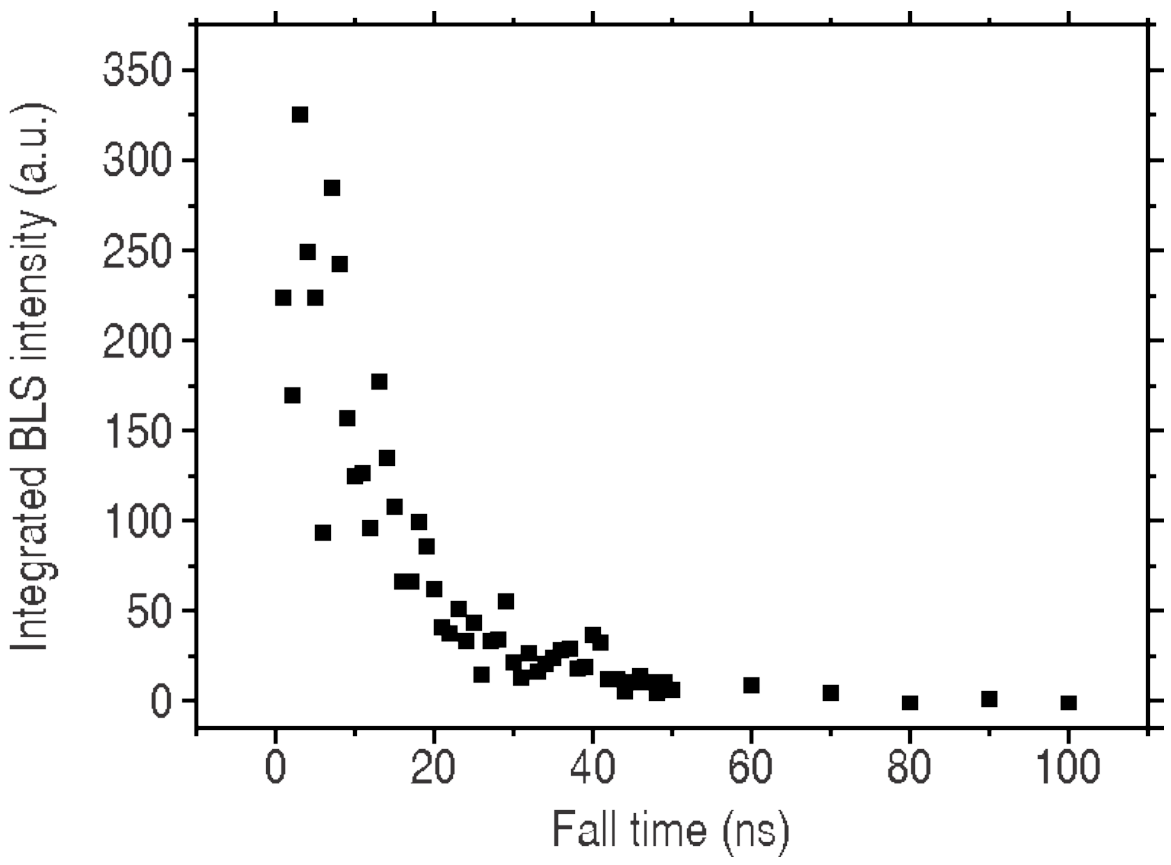

**Extended Data Fig. 10 | Dependence of the magnon BEC intensity on the fall time of the applied pulses.** Integrated BLS intensity of the fundamental mode as a function of the fall time of the applied current pulses. Only for fall times shorter than the lifetime of magnons $t_{\mathrm{m}} = 21 \pm 4$ ns a clear accumulation at the bottom of the spectrum is observed.

**Influence of thermally induced spin currents on the magnon BEC by rapid cooling.** An additional experiment using an Al (7 nm)/Au (5 nm) heating layer on top of a 2 µm wide and 34 nm thick YIG waveguide was performed to show that the main mechanism underlaying the BEC is the Rapid Cooling. Owing a small spin-orbit interaction in Al, a negligible spin current can be thermally injected into YIG from the heating layer by the spin Seebeck effect. The length of the heating area of 4 µm was the same as in the main experiment. The magnetic field of 188 mT was applied in-plane perpendicular to the waveguide long axis. Extended Data Figure 11 shows the accumulation of magnons after the 50 ns long current pulse applied. A voltage of 3.1 V is applied during this time, whereas the resistance of the structure under investigation is around 2 kOhm. The clear peak in the integrated BLS intensity (see panel b) is observed at the bottom frequency of the fundamental mode after the current pulse is switched off. This experiment directly confirms that the rapid cooling rather than spin-orbit mediated Spin Seebeck effect[17] is responsible for the reported BEC of magnons.

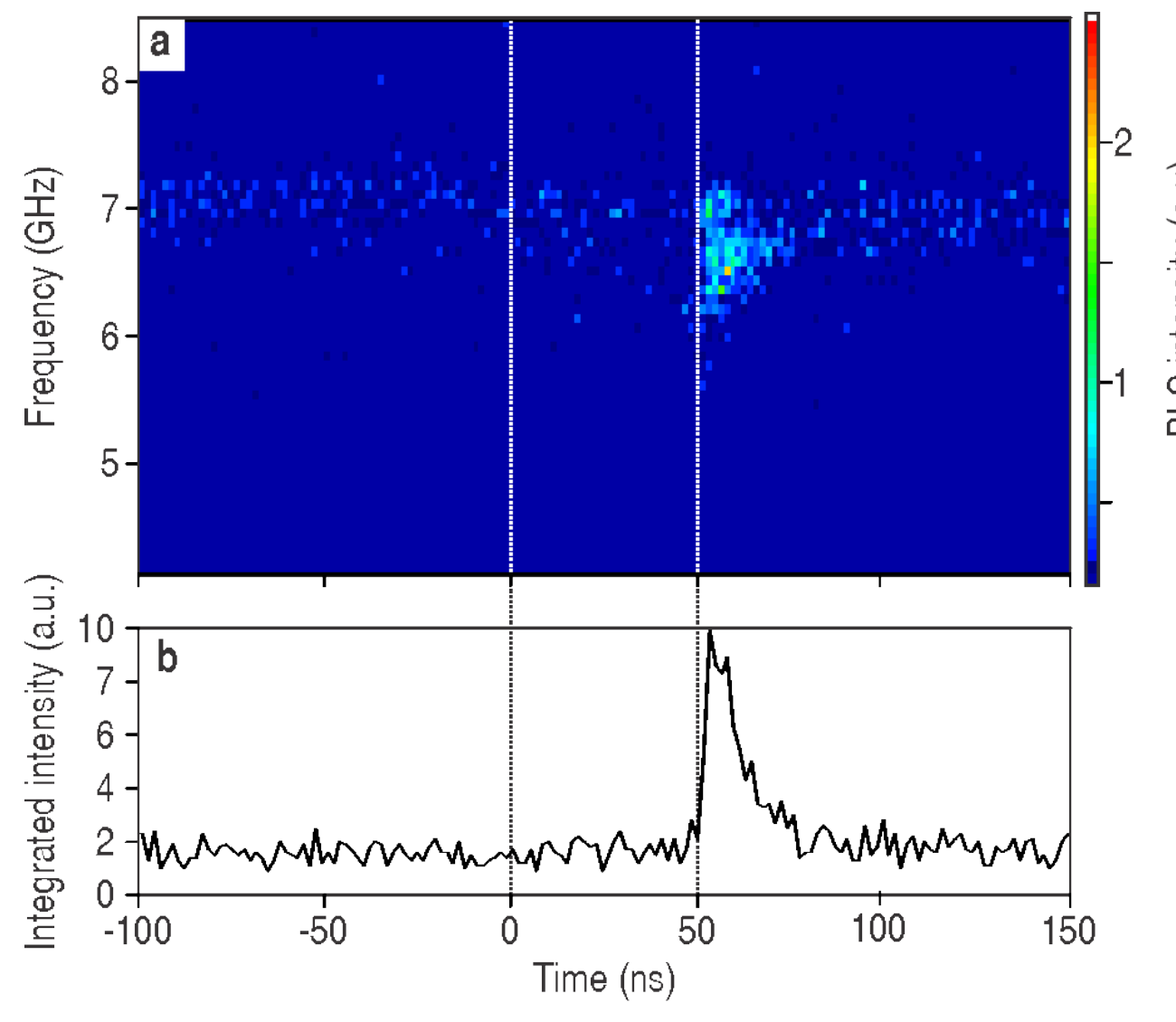

**Extended Data Fig. 11| Rapid cooling BEC generated in an Au/Al/YIG structure. a,** Time resolved BLS spectrum for the case when a 50 ns long pulse is applied **b,** Integrated BLS intensity over the frequency range shown in **a**.